\documentclass[lettersize,journal,referee]{IEEEtran}
\usepackage{bm}
\usepackage{amsmath,amsfonts}
\usepackage{algorithmic}
\usepackage{algorithm}
\usepackage{array}
\usepackage{textcomp}
\usepackage{stfloats}
\usepackage{url}
\usepackage{verbatim}
\usepackage{graphicx}
\usepackage{cite}
\graphicspath{{./Figure1/}}
\usepackage{subfig}
\usepackage{setspace}
\usepackage{amsmath}
\usepackage{indentfirst}
\usepackage{subfig}
\usepackage[dvipsnames]{xcolor}
\usepackage{amsmath}
\usepackage{booktabs}

\begin{document}
\title{Conditional Denoising Diffusion Probabilistic Model for Ground-roll Attenuation}
\author{Yuanyuan~Li, Hao~Zhang,  Jianping~Huang and Zhenchun~Li

\thanks{ 
This study is supported by the Marine S\&T Fund of Shandong Province for Pilot National Laboratory for Marine Science and Technology (Qingdao) (no. 2021QNLM020001), the Major Scientific and Technological Projects of Shandong Energy Group (no. SNKJ2022A06-R23), the Funds of Creative Research Groups of China (no. 41821002), National Natural Science Foundation of China Outstanding Youth Science Fund Project (Overseas) (no. ZX20230152), the Major Scientific and Technological Projects of CNPC (no. ZD2019-183-003). (\textit{Corresponding author: Jianping Huang})

Yuanyuan Li, Hao Zhang, Jianping Huang are with the Department of Geophysics, China University of Petroleum, Qingdao 266580, China (e-mail: yuanyuanli@upc.edu.cn; zhanghao@s.upc.edu.cn; jphuang@upc.edu.cn). 

}
}
\markboth{\tiny This work has been submitted to the IEEE for possible publication. Copyright may be transferred without notice, after which this version may no longer be accessible.}%
{Shell \MakeLowercase{\textit{et al.}}: Bare Demo of IEEEtran.cls for IEEE Journals}
\maketitle


\begin{abstract}
Ground-roll attenuation is a challenging seismic processing task in land seismic survey. The ground-roll coherent noise with low frequency and high amplitude  seriously contaminate the valuable reflection events, corrupting the quality of seismic data. The transform-based filtering methods leverage the distinct characteristics of the ground roll and seismic reflections within the transform domain to attenuate the ground-roll noise. However, the ground roll and seismic reflections often share overlaps in the transform domain, making it challenging to remove ground-roll noise without also attenuating useful reflections. We propose to apply a conditional diffusion denoising probabilistic model (c-DDPM) to attenuate the ground-roll noise and recover the reflections efficiently. We prepare the training dataset by using the finite-difference modelling method and the convolution modelling method. After the training process, the c-DDPM can generate the clean data given the seismic data as condition. The ground roll obtained by subtracting the clean data from the seismic data might contain some residual reflection energy. Thus, we further improve the c-DDPM to allow for generating the clean data and ground roll simultaneously. We then demonstrate the feasibility and effectiveness of our proposed method by using the synthetic data and the field data. The methods based on the local time-frequency (LTF) transform and U-Net are also applied to these two examples for comparing with our proposed method. The test results show that the proposed method perform better in attenuating the ground-roll noise from the seismic data than the LTF and U-Net methods.  

\end{abstract}

\begin{IEEEkeywords}
conditional denoising diffusion probabilistic model, ground-roll attenuation, deep generative model. 
\end{IEEEkeywords}



\section{Introduction}
Ground roll is often dominated by Rayleigh surface waves, generated from seismic sources near the Earth's surface \cite{beresford1988dispersive}. In land seismic survey, ground roll is a typical coherent noise with the main characteristics of low frequency, low velocity, high amplitude \cite{saatcilar1988method,liu1999ground,porsani2009ground}. Dispersion is also observed in ground roll, that is, different frequency components propagate at different velocities and arrive at the receiver at different times \cite{al1981dispersion}. Thus, ground roll usually appears as fan-shaped distribution with downward oblique straight lines in seismic recording. The ground-roll noise will corrupt the useful reflections, and thus needing to be attenuated for improving the quality of seismic data and the fidelity of seismic reflection imaging \cite{yilmaz2001seismic,chen2015ground}.  

The transform-based filtering methods play an important role for suppressing ground roll in seismic data processing. Since ground roll typically has lower frequency than seismic reflection signals, low-cut filters or high-pass filters can be applied to remove the low-frequency noise associated with ground roll. However, they also remove the low-frequency components of seismic signals. Thus, a variety of transform-based seismic processing methods are developed to suppress ground-roll noise while preserving valuable seismic signals. These methods leverage the distinct characteristics of ground roll and seismic reflections within the transform domain, such as frequency-wavenumber (\(f-k\)) transform \cite{treitel1967some,beresford1989suppression}, local time-frequency (LTF) transform \cite{liu2013seismic}, Radon transform \cite{trad2003latest,henley2003coherent}, wavelet transform \cite{deighan1997ground,zhang2003physical,gilles2013empirical}, S-transform \cite{stockwell1996localization,askari2008ground} and Karhunen-Loeve (KL) transform \cite{jones1987signal,liu1999ground,montagne2006optimized}. These transforms provides a different representation of seismic data, enabling the identification and attenuation of ground roll in the transformed seismic data. However, the ground roll and reflections can have significant overlap in the transform domain in complex real data scenario. This will make it very challenging to separate ground roll from seismic reflections.

Deep learning (DL) has received widespread attention in the field of geophysics, involving various tasks in seismic processing, inversion, and interpretation \cite{wu2019faultseg3d,yu2021deep,song2021wavefield,li2021deep,mousavi2022deep,harsuko2022storseismic}. In recent years, the DL methods are increasingly applied for attenuating ground roll \cite{li2018deep,liu2020should,zhang2021ground,pham2022physics,yang2023deep,xing2024ground}. Li \textit{et al.} applied a Convolutional Neural Network (CNN) for learning the features of scattered ground-roll noise and then removing the learned noise from shot gathers \cite{li2018deep}. Kaur \textit{et al.} applied the LTF and regularized non-stationary regression to a few shot gathers to prepare the labeled data for training a Generative Adversarial Network (GAN) for ground-roll attenuation \cite{kaur2020seismic}. Yuan \textit{et al.} also adopted the GAN to attenuate ground roll in seismic data \cite{yuan2020ground}. Oliveira \textit{et al.} combined a CNN and a conditional GAN to build the self-supervised two-step scheme for attenuating ground-roll noise \cite{oliveira2020self}. To avoid the preparation for clean data as training labels, Liu \textit{et al.} designed the  blind-fan networks to suppress the coherent ground-roll noise in seismic data in a self-supervised manner, but a careful selection of the mask structure is really needed \cite{liu2023self}.

Denoising Diffusion Probabilistic Model (DDPM) is an advanced generative model with superior ability to generate high-quality, diverse and complex samples \cite{ho2020denoising}. Considering the powerful performance of DDPM, Durall \textit{et al.} introduced a DDPM to handle seismic data processing tasks, including demultiple, denoising and interpolation \cite{durall2023deep}. Conditional DDPM (c-DDPM) further extends the capability of standard DDPM by guiding the generation process with the condition \cite{ho2022classifier}. Thus, we propose to apply the c-DDPM to suppress ground-roll noise in seismic data. 

Typically, we can design the clean data as the target data distribution for the c-DDPM. Once the c-DDPM is trained, we can generate the clean data corresponding to the conditioning seismic data, contaminated by ground-roll noise. The ground roll can then be obtained by subtracting the generated clean data from the seismic data. However, the obtained ground roll usually contains some reflection residuals due to the inaccurate prediction of the clean data. Thus, we propose an improved c-DDPM to predict the ground roll with higher accuracy. Here, we adjust the forward and reverse processes of the diffusion model and modify the network architecture. Given the conditioning seismic data, the improved c-DDPM is able to generate the clean data and ground roll simultaneously. 

To the best of our knowledge, this is the first work that applies c-DDPM to ground-roll attenuation task, showing promising results compared with the methods based on LTF and U-Net. The rest of the paper is organized as follows. First, we introduce the theory of the conventional c-DDPM and our improved c-DDPM. Secondly, we introduce the network architecture and how to prepare the training dataset and train the model. In the Examples section, we use test the synthetic and field data to test our proposed method. The LTF and U-Net methods are also applied to these data for comparing with the proposed method. At the end, we make discussions and conclusion.

\section{C-DDPM}
\subsection{Conventional c-DDPM }   
For conventional c-DDPM, the target data distribution should be either clean data or ground roll data. Once one of them is obtained by the c-DDPM, the other one can be obtained by subtracting it from the noisy data. In the conventional c-DDPM, we typically set the clean data $\mathbf{x}_0$ as the target distribution and the noisy data $\mathbf{y}$ as the condition. Correspondingly, the ground roll $\mathbf{z}_0$ can be obtained by subtracting the generated clean data from the noisy data. The forward process for the c-DDPM is defined on a Markov chain within $T$ timesteps:
\begin{equation} \label{eq:1}
\begin{aligned}
q(\mathbf{x}_{1:T}|\mathbf{x}_0)=\prod\limits_{t=1}^T q(\mathbf{x}_t|\mathbf{x}_{t-1}),
\end{aligned}
\end{equation}
where $\mathbf{x}_0$ is sampled from the training datasets $ q(\mathbf{x}_0|\mathbf{y})$, $\mathbf{x}_1,..,\mathbf{x}_T$ are the latent variables fixed on the Markov chain. $q(\mathbf{x}_t|\mathbf{x}_{t-1})$ is the pre-defined Gaussian translation and obtains $\mathbf{x}_t$ by adding random noise to $\mathbf{x}_{t-1}$:
\begin{equation} \label{eq:2}
\begin{aligned}
q(\mathbf{x}_t|\mathbf{x}_{t-1})=\mathcal{N}(\mathbf{x}_t;\sqrt{1-\beta_t}\mathbf{x}_{t-1},\beta_t\mathbf{I}),
\end{aligned}
\end{equation}
where $\beta_t \in (0,1)$ is a predefined variance schedule that increases with timestep $t$.
 The property of the Markov chain enables:
\begin{equation} \label{eq:3}
\begin{aligned}
q(\mathbf{x}_t|\mathbf{x}_0)=\mathcal{N}(\mathbf{x}_t;\sqrt{\bar{\alpha}_t}\mathbf{x}_0,(1-\bar{\alpha}_t)\mathbf{I}).
\end{aligned}
\end{equation}
A more explicit form of Eq. \ref{eq:3} is:
\begin{equation} \label{eq:4}
\begin{aligned}
\mathbf{x}_t=\sqrt{\bar{\alpha}_t}\mathbf{x}_0+\sqrt{1-\bar{\alpha}_t}\boldsymbol{\epsilon_{\mathbf{x}}},\quad\boldsymbol{\epsilon_{\mathbf{x}}}\sim\mathcal{N}(\mathbf{0},\mathbf{I}),
\end{aligned}
\end{equation}
where ${\alpha_{t}}=1-\beta_{t}$ and ${\bar{\alpha}}_{t}=\prod_{s=1}^{t}{\alpha}_{s}$.
At timestep $T$, $\mathbf{x}_t$ converges to a prior distribution, i.e, a standard normal distribution. Thus, the forward process is a diffusion process that gradually converts a clean image to a pure noise image.

The reverse process is also defined on a Markov chain, which converts pure noise 
$\mathbf{x}_T\sim\mathcal{N}(\mathbf{0},\mathbf{I})$ to the data distribution $\mathbf{x_0}$, under the guidance of $\mathbf{y}$:
\begin{equation} \label{eq:5}
\begin{aligned}
p_\theta(\mathbf{x}_{0:T}|y)=p(\mathbf{x}_T)\prod\limits_{t=1}^Tp_\theta(\mathbf{x}_{t-1}|\mathbf{x}_t,\mathbf{y}),
\end{aligned}
\end{equation}
where the Gaussian translation $p_\theta(\mathbf{x}_{t-1}|\mathbf{x}_t,\mathbf{y})$ has learned mean $\bm{\mu}_{\theta}$ and fixed variance $\mathbf{\Sigma}_\theta={\sigma_t}^2$:

\begin{equation} \label{eq:6}
\begin{aligned}
p_\theta(\mathbf{x}_{t-1}|\mathbf{x}_t,y)=\mathcal{N}(\mathbf{x}_{t-1};\bm{\mu}_{\theta}\left(\mathbf{x}_{t},\mathbf{y},t\right),\mathbf{\Sigma}_\theta(\mathbf{x}_t,\mathbf{y},t)).
\end{aligned}
\end{equation}
For the detailed inference of the formulas, see \cite{ho2020denoising}. The optimization function of the network is:

\begin{equation}  \label{eq:7}
\begin{aligned}
\mathbb{E}_{t,\mathbf{x}_t,\boldsymbol{\epsilon}}\Big[\left|\boldsymbol{\epsilon}-\boldsymbol{\epsilon}_{\theta}(\mathbf{x}_{t},\mathbf{y},t)\right|\big].
\end{aligned}
\end{equation} 
 After training the network, we can gradually generate the clean data $\hat{\mathbf{x}_0}$ with the sampling step in c-DDPM:
\begin{equation}  \label{eq:8}
\begin{split}
\hat{\mathbf{x}}_{t-1}=\frac{1}{\sqrt{\alpha_t}}\left(\hat{\mathbf{x}}_t-\frac{1-\alpha_t}{\sqrt{1-\bar{\alpha}_t}}{\boldsymbol{\epsilon}}_{\theta}(\hat{\mathbf{x}}_{t},\mathbf{y},t)\right)+\sigma_t\mathbf{\epsilon_{\mathbf{x}}},
\end{split}
\end{equation} 
where $\hat{\mathbf{x}}_{t}$ is the estimated latent variable in the sampling process.
\begin{figure*} [htp!]
	\centering
 \includegraphics[width=2\columnwidth]{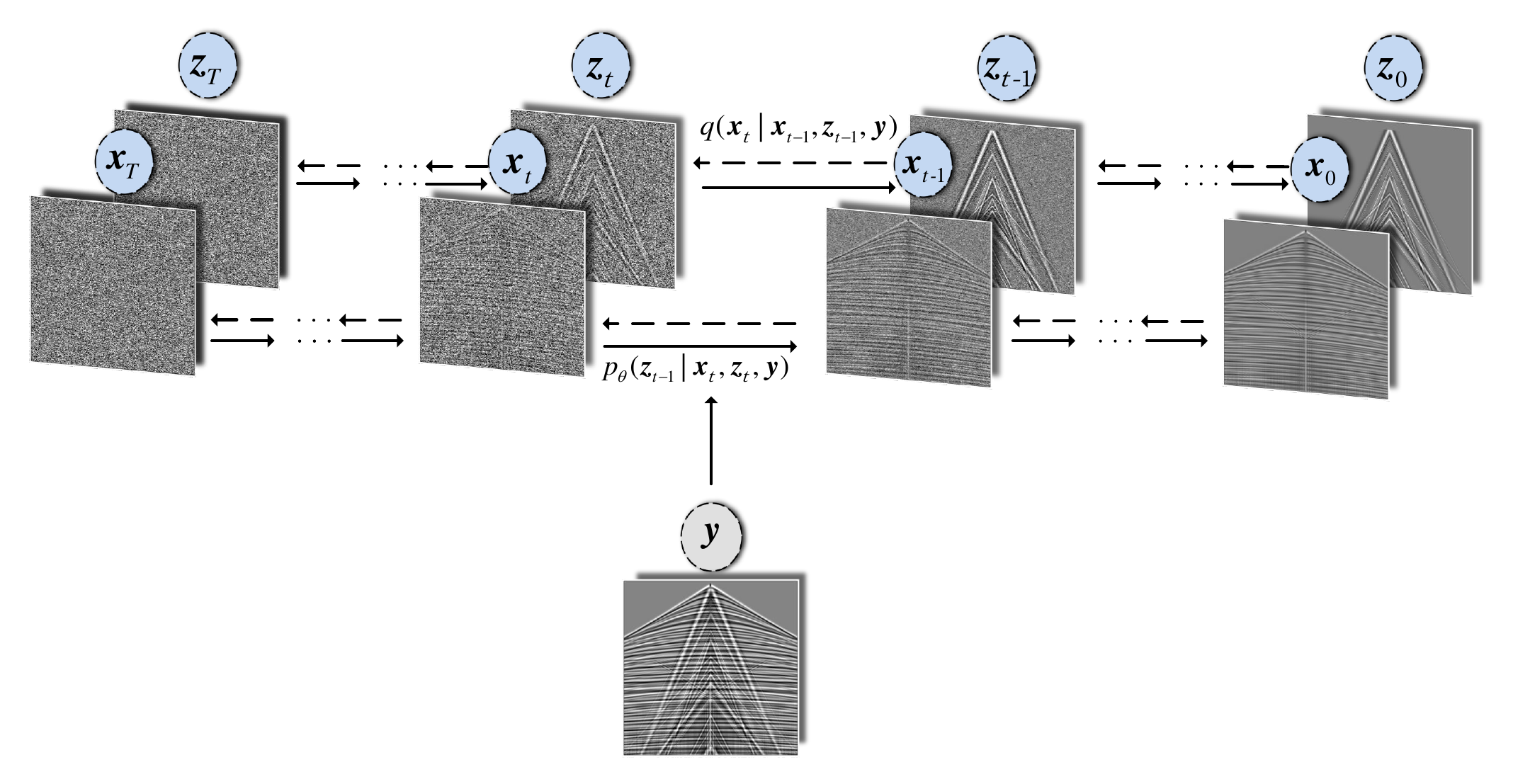}
 \caption{An illustration of the forward and reverse process in the improved c-DDPM. The improved c-DDPM sets the clean data and ground roll data as two target data distributions. The dashed line indicates the forward process, which gradually converts a clean image to a pure noise image. The solid line indicates the reverse process, which gradually converts a pure noise image to the clean image. }
 \label{fig:cddpm}
 \end{figure*} 
 
\subsection{Improved c-DDPM} 
 One limit of the conventional c-DDPM is that the ground roll $\hat{\mathbf{z}}_0$ obtained by subtracting the predicted clean data $\hat{\mathbf{x}}_0$ from the noisy data $\mathbf{y}$ usually suffer from some leakage of reflections because of the inaccurate prediction of $\hat{\mathbf{x}}_0$. Thus, we design two target distribution, clean data and ground roll, in the improved c-DDPM given the noisy data as condition. The forward and reverse processes for the improved c-DDPM are shown in Fig. \ref{fig:cddpm}. In detail, the forward process in the improved c-DDPM is
\begin{equation} \label{eq:9}
\begin{aligned}
q(\mathbf{x}_{1:T}|\mathbf{x}_0)=\prod\limits_{t=1}^T q(\mathbf{x}_t|\mathbf{x}_{t-1}),
\end{aligned}
\end{equation}
\begin{equation} \label{eq:10}
\begin{aligned}
q(\mathbf{z}_{1:T}|\mathbf{z}_0)=\prod\limits_{t=1}^T q(\mathbf{z}_t|\mathbf{z}_{t-1}),
\end{aligned}
\end{equation}
where $\mathbf{x}_0$ and $\mathbf{z}_0$ refer to the clean data and ground roll, respectively.  
The reverse process is formulated as: 
\begin{equation} \label{eq:11}
\begin{aligned}
p_\theta(\mathbf{\hat{x}}_{0:T}|\mathbf{\hat{z}}_{0:T},\mathbf{y})=p(\mathbf{\hat{x}}_T)\prod\limits_{t=1}^Tp_\theta(\mathbf{\hat{x}}_{t-1}|\mathbf{\hat{x}}_t,\mathbf{\hat{z}}_t,\mathbf{y}),
\end{aligned}
\end{equation}
\begin{equation} \label{eq:12}
\begin{aligned}
p_\theta(\mathbf{\hat{z}}_{0:T}|\mathbf{\hat{x}}_{0:T},\mathbf{y})=p(\mathbf{\hat{z}}_T)\prod\limits_{t=1}^Tp_\theta(\mathbf{\hat{z}}_{t-1}|\mathbf{\hat{x}}_t,\mathbf{\hat{z}}_t,\mathbf{y}).
\end{aligned}
\end{equation}
Correspondingly, the optimization function becomes
\begin{equation}  \label{eq:13}
\begin{aligned}
\mathrm{E}_{t,\mathbf{x}_t,\mathbf{z}_t,\mathbf{y},\boldsymbol{\epsilon}}\big[|\boldsymbol{\epsilon}_{\mathbf{x}}-\boldsymbol{\epsilon}_{\theta,\mathbf{x}}(\mathbf{x}_{t},\mathbf{z}_{t},\mathbf{y},t)|
+|\boldsymbol{\epsilon}_{\mathbf{z}}-\boldsymbol{\epsilon}_{\theta,\mathbf{z}}(\mathbf{x}_{t},\mathbf{z}_{t},\mathbf{y},t)|\big],
\end{aligned}
\end{equation}
where $\boldsymbol{\epsilon}_{\theta,\mathbf{x}}(\mathbf{x}_{t},\mathbf{z}_{t},\mathbf{y},t)$ and $\boldsymbol{\epsilon}_{\theta,\mathbf{z}}(\mathbf{x}_{t},\mathbf{z}_{t},\mathbf{y},t)$ represents the estimated noise for sampling $\mathbf{x}_{t-1}$ and $\mathbf{z}_{t-1}$ respectively. Note that, both the noise are jointly estimated from $\mathbf{x}_{t}$ and $\mathbf{z}_{t}$ with the network.
Once the training process is finished, we can gradually generate the clean data $\hat{\mathbf{x}}_{t}$ and the ground roll $\hat{\mathbf{z}}_{t}$ with the noisy data as condition:
\begin{equation} \label{eq:14}
\begin{split}
\hat{\mathbf{x}}_{t-1}&=\frac{1}{\sqrt{\alpha_t}}\left(\mathbf{\hat{x}}_t-\frac{1-\alpha_t}{\sqrt{1-\bar{\alpha}_t}}\boldsymbol{\epsilon}_{\theta,\mathbf{x}}(\mathbf{\hat{x}}_t,\mathbf{\hat{z}}_t,\mathbf{y},t)\right)\\
                       &+\sigma_t{\boldsymbol{\epsilon}_\mathbf{x}}, {\boldsymbol{\epsilon}_\mathbf{x}}\sim\mathcal{N}(\mathbf{0},\mathbf{I}),
\end{split}
\end{equation}
\begin{equation} \label{eq:15}
\begin{split}
\hat{\mathbf{z}}_{t-1}&=\frac{1}{\sqrt{\alpha_t}}\left(\mathbf{\hat{z}}_t-\frac{1-\alpha_t}{\sqrt{1-\bar{\alpha}_t}}\boldsymbol{\epsilon}_{\theta,\mathbf{x}}(\mathbf{\hat{x}}_t,\mathbf{\hat{z}}_t,\mathbf{y},t)\right)\\
&+\sigma_t{\boldsymbol{\epsilon}_\mathbf{z}},  {\boldsymbol{\epsilon}_\mathbf{z}}\sim\mathcal{N}(\mathbf{0},\mathbf{I}).
\end{split}
\end{equation}

The proposed training process and sampling process of the improved c-DDPM are shown in Algorithm \ref{alg:1} and Algorithm \ref{alg:2}, respectively.

\begin{algorithm}
        \renewcommand{\algorithmicrequire}{\textbf{Input:}}
        \caption{Training  of the improved c-DDPM}
        \label{alg:1}
        \begin{algorithmic}[1]
            \REQUIRE prepared dataset $q(\mathbf{x}_0,\mathbf{z}_0,\mathbf{y})$ ($\mathbf{x}_0$ and $\mathbf{z}_0$ is the clean data and the ground roll data respectively. $\mathbf{y}$ is  the noisy input), diffusion time steps $T$
		\WHILE{not converged}
		\STATE sample $\mathbf{x}_0$, $\mathbf{z}_0$, $\mathbf{y}$ from $q(\mathbf{x}_0,\mathbf{z}_0,\mathbf{y})$
		\STATE sample $t$ from Uniform(\{$1,...,T$\})	
            \STATE sample $\boldsymbol{\epsilon}_{\mathbf{x}}$, $\boldsymbol{\epsilon}_{\mathbf{z}}$ from $\mathcal{N}(\mathbf{0},\mathbf{I})$,  
            \STATE $\mathbf{z}_t=\sqrt{\bar{\alpha}_t}\mathbf{z}_0+\sqrt{1-\bar{\alpha}_t}\boldsymbol{\epsilon}_{\mathbf{x}}$
            \\$\mathbf{x}_t=\sqrt{\bar{\alpha}_t}\mathbf{x}_0+\sqrt{1-\bar{\alpha}_t}\boldsymbol{\epsilon}_{\mathbf{z}}$
            \STATE optimise $\mathrm{E}_{t,\mathbf{x}_t,\mathbf{z}_t,\mathbf{y},\boldsymbol{\epsilon}}\big[|\boldsymbol{\epsilon}_{\mathbf{x}}-\boldsymbol{\epsilon}_{\theta,\mathbf{x}}(\mathbf{x}_{t},\mathbf{z}_{t},\mathbf{y},t)|
+|\boldsymbol{\epsilon}_{\mathbf{z}}-\boldsymbol{\epsilon}_{\theta,\mathbf{z}}(\mathbf{x}_{t},\mathbf{z}_{t},\mathbf{y},t)|\big]$ in network $\theta$
        \ENDWHILE
	\end{algorithmic} 
\end{algorithm}

\begin{algorithm}[!ht]
    \renewcommand{\algorithmicrequire}{\textbf{Input:}}
    \renewcommand{\algorithmicensure}{\textbf{Output:}}
    \caption{Sampling of the improved c-DDPM}
    \label{alg:2}
    \begin{algorithmic}[1]
        \REQUIRE $\mathbf{y}$, $T$
        \ENSURE Clean data $\mathbf{\hat{x}}_0$ and Ground roll data $\mathbf{\hat{z}}_0$
        \STATE Sample $\mathbf{x}_T,\mathbf{z}_T$ from $\mathcal{N}(\mathbf{0},\mathbf{I})$
        \FOR{$t=T,...,1$}
            \IF{$t > 1$}
                \STATE sample $\mathbf{\epsilon_x}$, $\mathbf{\epsilon_z}$ from $\mathcal{N}(\mathbf{0},\mathbf{I})$
            \ELSE
                \STATE $\mathbf{\epsilon_x}=\mathbf{\epsilon_z}=0$  
            \ENDIF
            \STATE predict $\boldsymbol{\epsilon}_{\theta,\mathbf{x}}(\mathbf{x}_{t},\mathbf{y},t)$,$\boldsymbol{\epsilon}_{\theta,\mathbf{z}}(\mathbf{x}_{t},\mathbf{y},t)$ through trained network $\theta$
            \STATE $\hat{\mathbf{x}}_{t-1}=\frac{1}{\sqrt{\alpha_t}}\left(\mathbf{\hat{x}}_t-\frac{1-\alpha_t}{\sqrt{1-\bar{\alpha}_t}}\boldsymbol{\epsilon}_{\theta,\mathbf{x}}(\mathbf{\hat{x}}_t,\mathbf{\hat{z}}_t,t)\right)+\sigma_t\mathbf{\epsilon_z}$\\
            $\hat{\mathbf{z}}_{t-1}=\frac{1}{\sqrt{\alpha_t}}\left(\mathbf{\hat{z}}_t-\frac{1-\alpha_t}{\sqrt{1-\bar{\alpha}_t}}\boldsymbol{\epsilon}_{\theta,\mathbf{z}}(\mathbf{\hat{x}}_t,\mathbf{\hat{z}}_t,t)\right)+\sigma_t\mathbf{\epsilon_z}$
        \ENDFOR
    \end{algorithmic}
\end{algorithm}
\section{Network architecture and training}

\begin{figure} [htp!]
	\centering
 \subfloat[\label{fig:net-1c}]{
		\includegraphics[width=1\columnwidth]{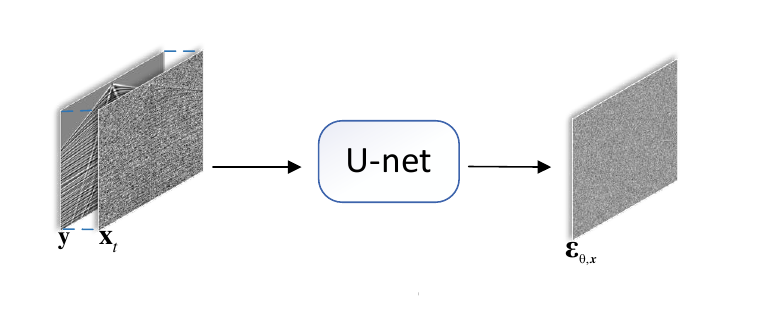}}	\\
   \subfloat[\label{fig:net-2c}]{
		\includegraphics[width=1\columnwidth]{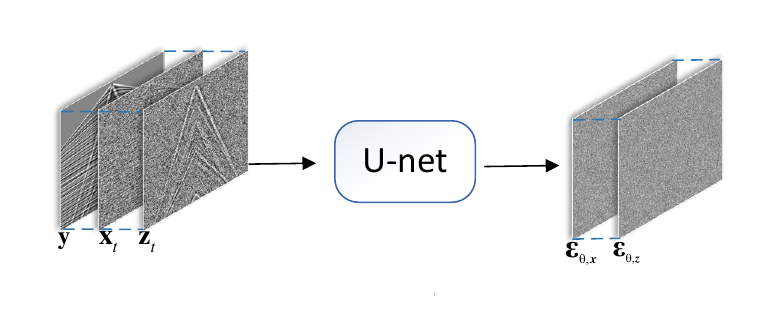}}

 \caption{A graphic illustration of the input layers and output layers for the conventional c-DDPM and improved c-DDPM. (a) The conventional c-DDPM. (b) The improved c-DDPM.}
 \end{figure}

\begin{figure} [htp!]
	\centering
		\includegraphics[width=1\columnwidth]{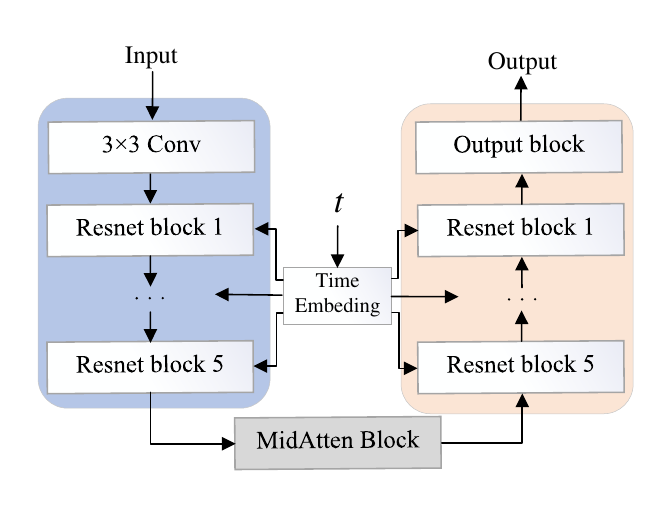}	
 \caption{A general illustration of the U-net used for the conventional c-DDPM and improved c-DDPM.}
 \label{fig:big-network}
 \end{figure}

We adopt the modified U-Net architecture, an adaptation of the network used by \cite{saharia2022image}, as the backbone for the c-DDPM. The only difference between the two U-Nets used for the conventional and improved c-DDPM is the channel numbers of the input and output layers. According to Eq. \ref{eq:7}, the U-Net used in the conventional c-DDPM takes the condition $\mathbf{y}$ and noisy clean data $\mathbf{x}_t$ as input and produces the estimated noise $\mathbf{\epsilon}_{\theta,\mathbf{x}}$ as output.  We concatenate $\mathbf{y}$ and $\mathbf{x}_t$ in the input layer with 2 channels. The input and output layer for the U-Net used in conventional c-DDPM is shown in Fig. \ref{fig:net-1c}. Compared with the U-Net in the conventional c-DDPM, we include the noisy ground roll data $\mathbf{z}_t$ as an additional input channel and the estimated noise $\mathbf{\epsilon}_{\theta,\mathbf{z}}$ as an additional output channel in the improved c-DDPM (Eq. \ref{eq:13}).
Fig. \ref{fig:net-2c} shows the input and output layers for the U-Net used in improved c-DDPM. For convenience, we refer to the conventional c-DDPM and the improved c-DDPM as "DDPM-1c" and "DDPM-2c", respectively. The architecture of the U-Net used jointly for the DDPM-1c and the DDPM-2c is shown in Fig. \ref{fig:big-network}. The U-Net consists of the encoder and decoder. The skip connection in the U-Net enables conveying the information from the encoder to the decoder. Our U-Net mainly includes 5 Resnet blocks in both the encoder and decoder, a MidAtten block that connects the encoder and decoder, a time-embedding block and an output block. The self-attention mechanism is introduced into the MidAtten block to help the U-Net to learn the features of the image\cite{vaswani2017attention}.Furthermore, the time embedding block introduces position encoding to inform the U-Net of the current time step in the reverse step \cite{vaswani2017attention}.\\

\begin{figure} [htp!]
	\centering
 \subfloat[\label{fig:zerooffset_simple_a}]{
		\includegraphics[width=0.3\columnwidth]{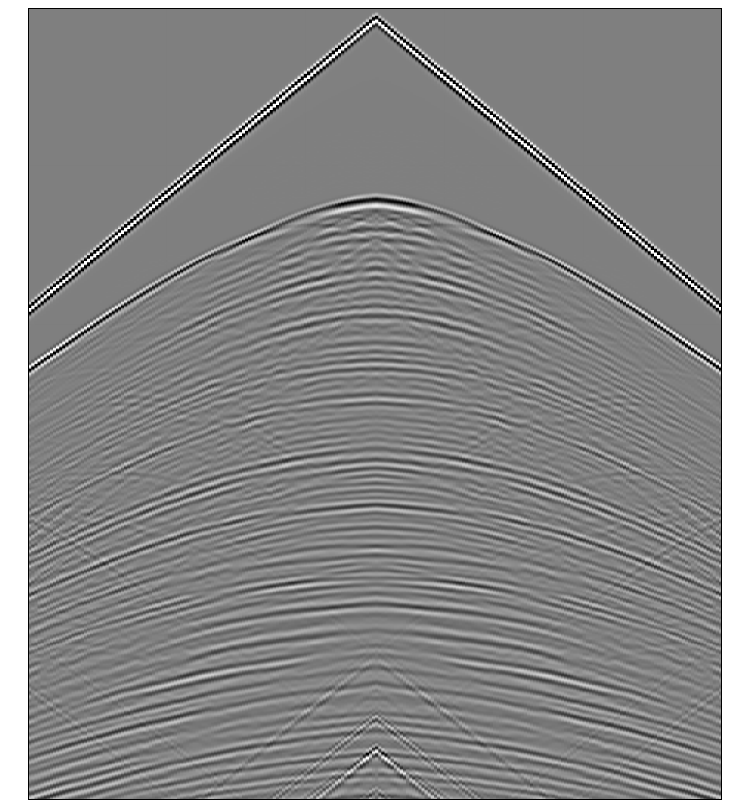}}	
   \subfloat[\label{fig:zerooffset_simple_b}]{
		\includegraphics[width=0.3\columnwidth]{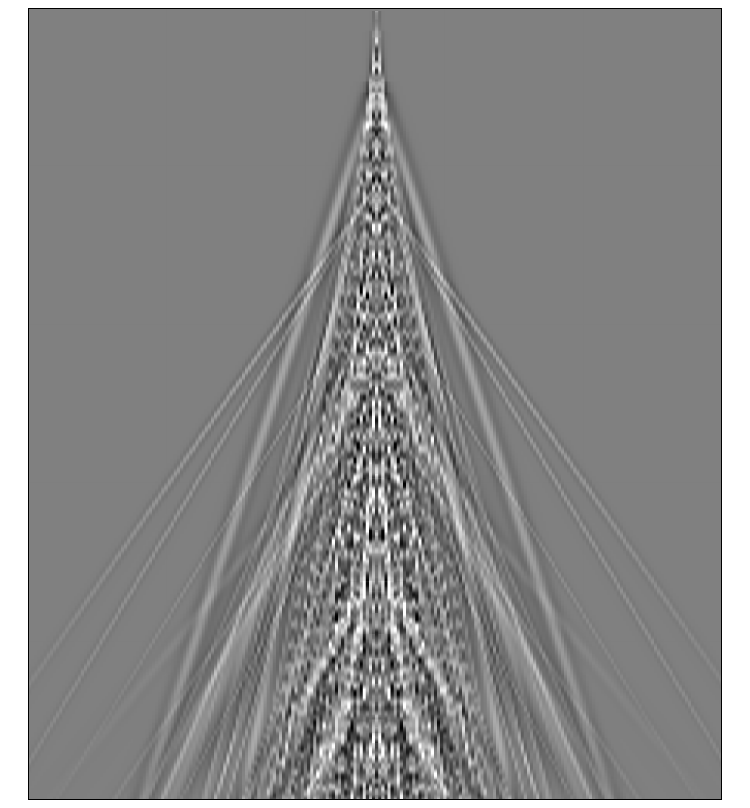}}
  \subfloat[\label{fig:zerooffset_simple_c}]{
		\includegraphics[width=0.3\columnwidth]{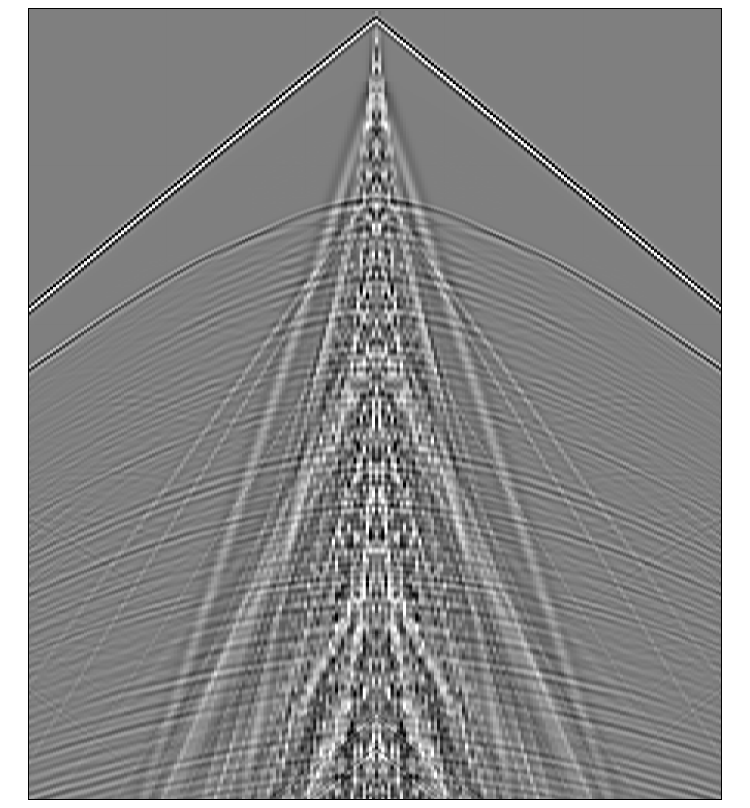}}\\
 \subfloat[\label{fig:zerooffset_simple_a}]{
		\includegraphics[width=0.3\columnwidth]{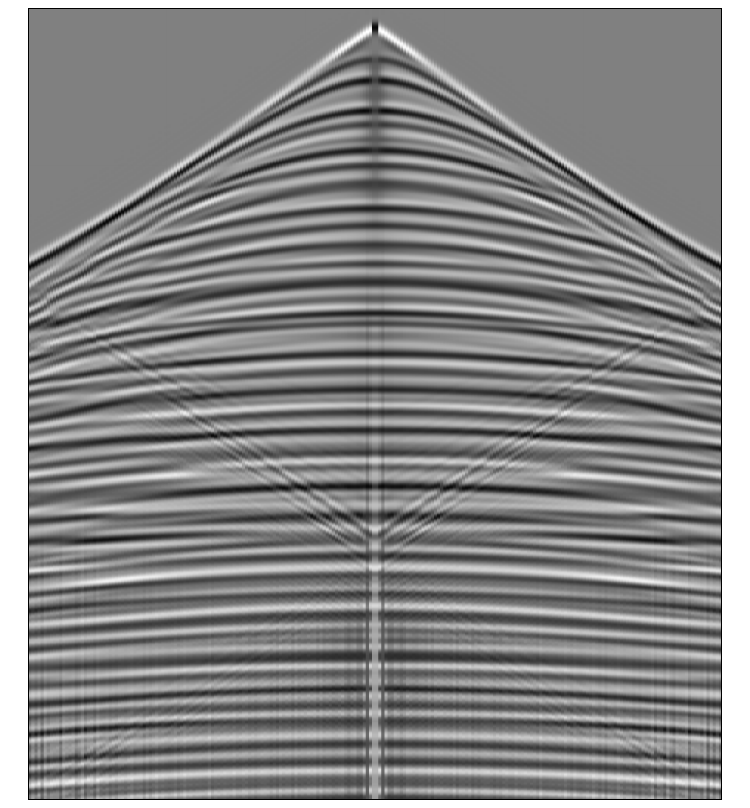}}	
   \subfloat[\label{fig:zerooffset_simple_b}]{
		\includegraphics[width=0.3\columnwidth]{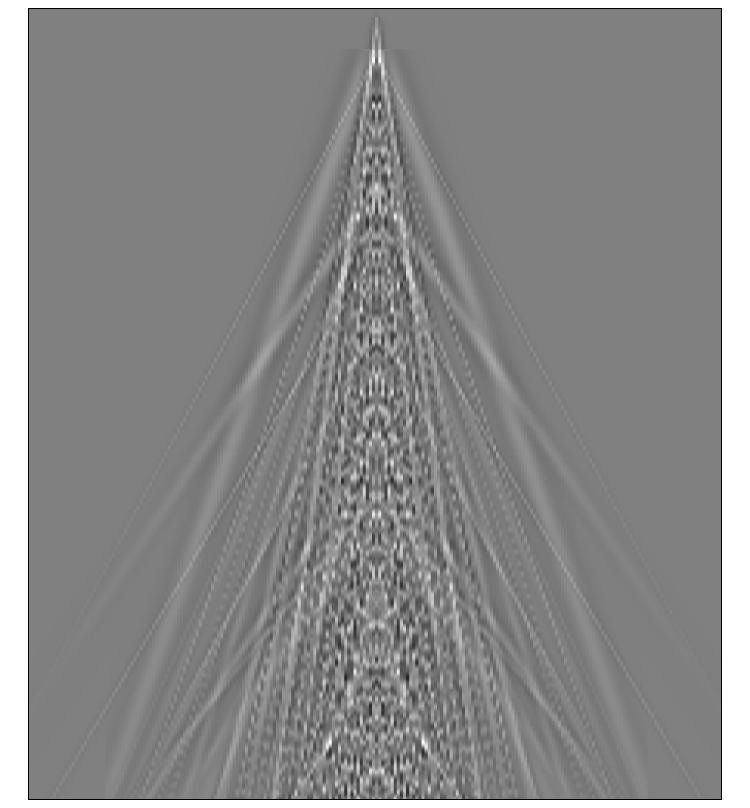}}
  \subfloat[\label{fig:zerooffset_simple_c}]{
		\includegraphics[width=0.3\columnwidth]{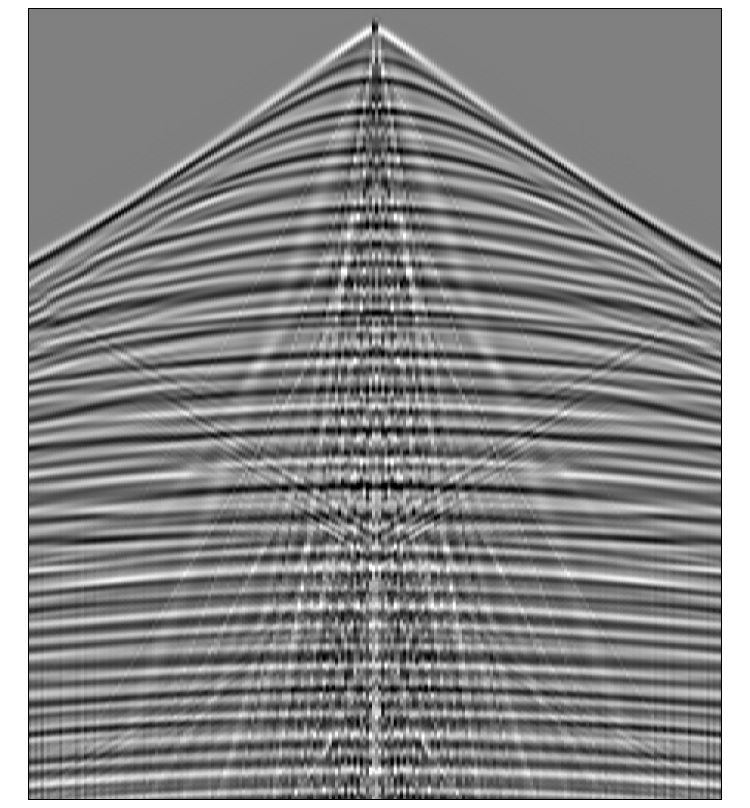}}\\
  \caption{Examples of the training datasets. (a) and (d) are two examples of  seismic clean data using finite difference modelling. (b) and (c) are the ground roll data using covolution modelling. (c) and (f) are the noisy data, after a summation of (a) and (b), (d) and (c), respectively}
    \label{fig:training datasets}
\end{figure}

Given the network architecture, we then prepare the training dataset for training the network. In  a supervised manner, the training dataset should include noisy seismic data as input and corresponding clean data as label. There are various methods, such as finite difference modeling and convolution modeling, for preparing the datasets.

Finite difference modeling is capable of generating seismic data with excellent dynamic features, but it is computationally intense. Convolution modeling is computationally simple and efficient and allows for flexible adjustment of the location, amplitude, and frequency of seismic events. However, the convolution modeling often fails to capture the dynamic features of seismic data. To balance the cost and accuracy of the data generation, we apply the finite difference modelling to produce the clean data accurately, while using the convolution modelling to generate coherent ground-roll noise efficiently.
Similar strategy is also used in \cite{yuan2020ground}. 

To prepare the clean data for training, we first simulate 792 shot gathers by using various layered velocity models and then extract 792 shot gathers from the 2007 BP Anisotropic Velocity Benchmark datasets. We generate the ground roll randomly by adjusting the location of linear reflectivity and the frequency and amplitude of the wavelet in the convolution modeling. To adapt the trained network to the data of the example section, we need to select a reasonable frequency range in the convolution modelling. When the ground roll are much stronger than the clean data, the performance of the network greatly deteriorates. Thus, we set the average amplitude of the ground roll to be 1.5-3 times larger than the clean data. We then make the summation of the clean data and ground roll as the noisy data. The dimension of each sample of the training datasets is fixed to 640 in the time direction and 224 in the horizontal. The volume of our training datasets is 1584. Two examples of the prepared datasets are shown in Figure \ref{fig:training datasets}.

We train the DDPM-1c and DDPM-2c with 63 epochs and the batch size of 4. We use an Adam optimization with an initial learning rate of 0.0001. As a comparison, we use the same training parameters to train the U-Net in a supervised fashion, where the noisy and clean data are used as the input and output of this U-Net, respectively. Then, we directly use the trained models in the synthetic and field data examples. 
\section{Examples}

\begin{figure*} [htp!]
	\centering
 \subfloat[\label{fig:aridm_a}]{
		\includegraphics[width=0.65\columnwidth]{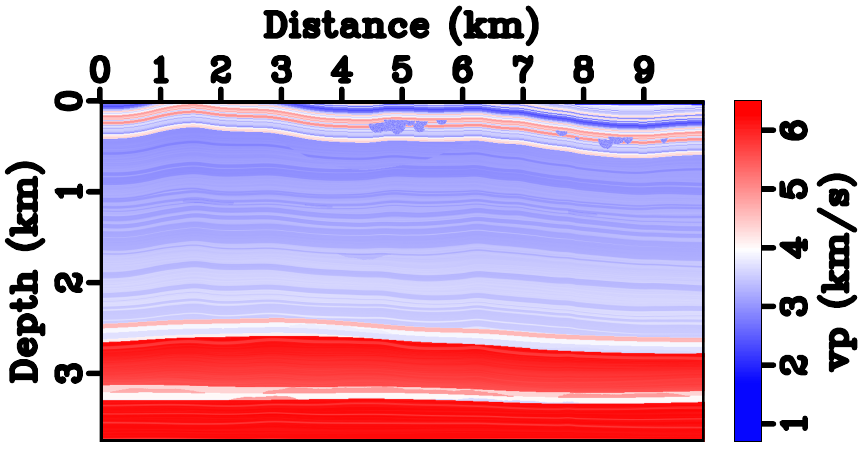}}	
   \subfloat[\label{fig:aridm_b}]{
		\includegraphics[width=0.65\columnwidth]{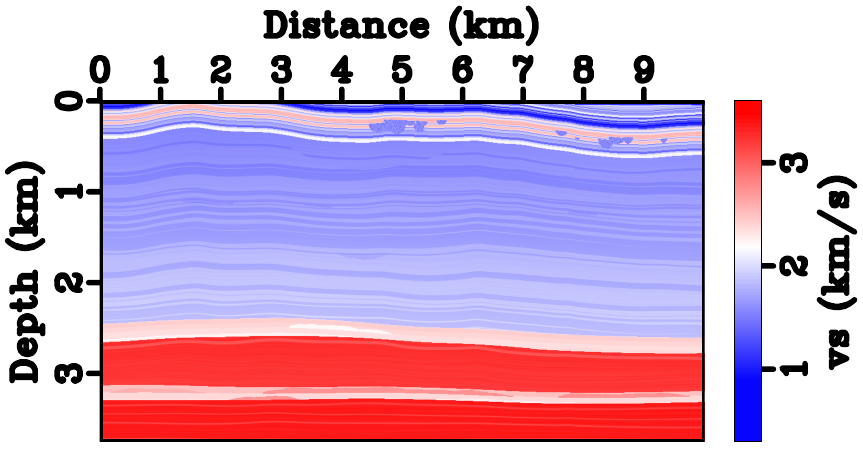}}
  \subfloat[\label{fig:aridm_c}]{
		\includegraphics[width=0.65\columnwidth]{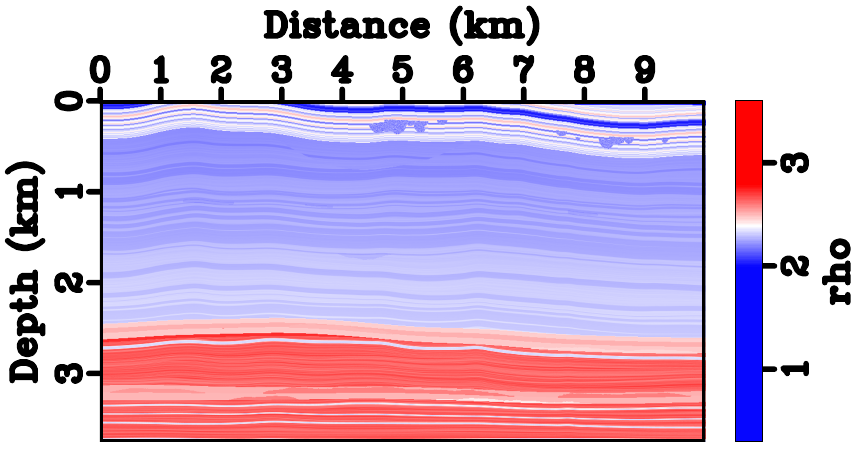}}
  \caption{A 2D slice of SEAM Arid model: (a)\(v_p\), (b)\(v_s\) and (c)\(\rho\).}
    \label{fig:aridm}
\end{figure*}

\begin{figure*} [htp!]
	\centering
 \subfloat[\label{fig:arid_a}]{
		\includegraphics[width=0.5\columnwidth]{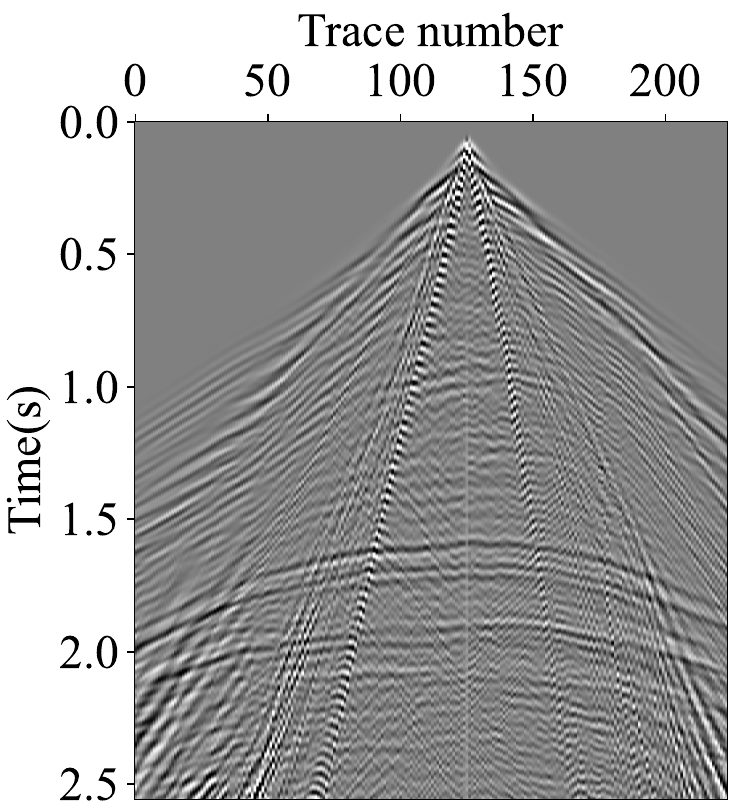}}	
   \subfloat[\label{fig:arid_b}]{
		\includegraphics[width=0.5\columnwidth]{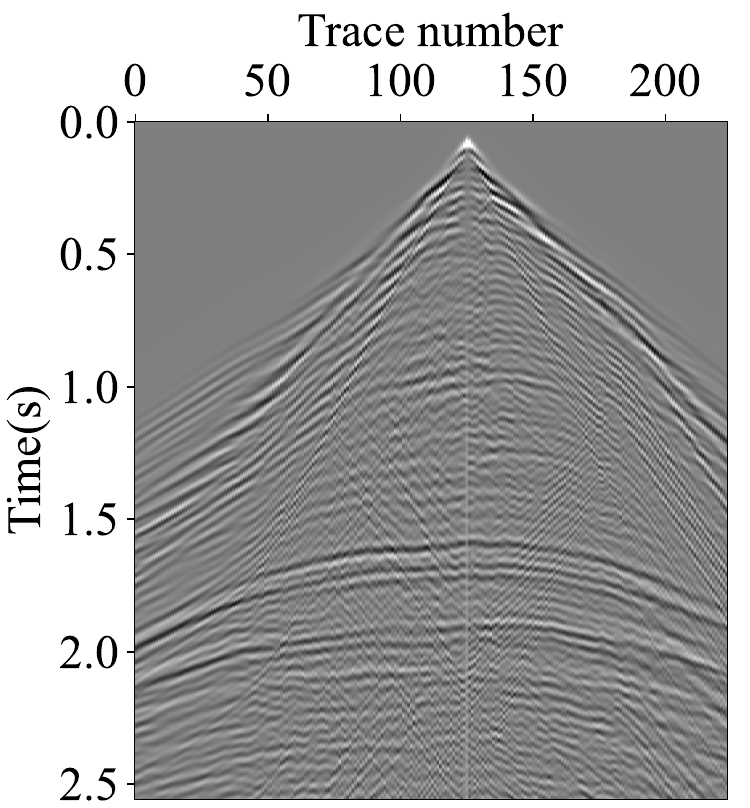}}
  \subfloat[\label{fig:arid_c}]{
		\includegraphics[width=0.5\columnwidth]{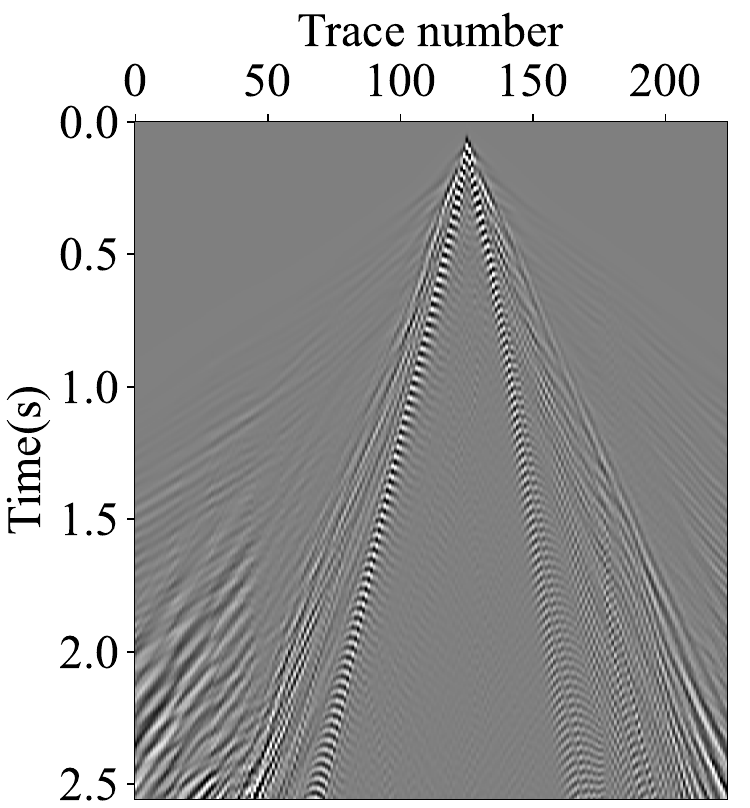}}\\
   \subfloat[\label{fig:arid_d}]{
		\includegraphics[width=0.5\columnwidth]{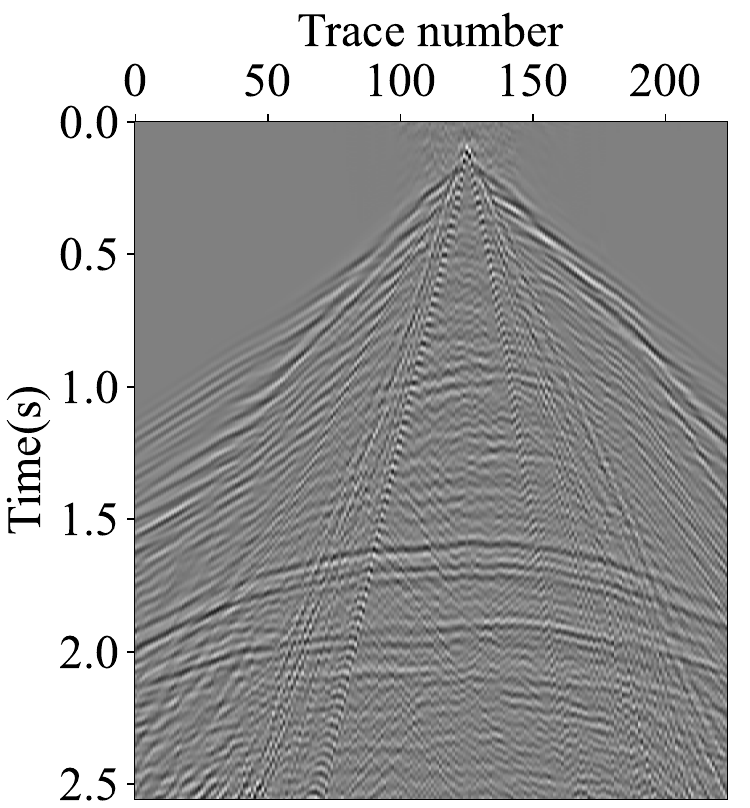}}	
   \subfloat[\label{fig:arid_e}]{
		\includegraphics[width=0.5\columnwidth]{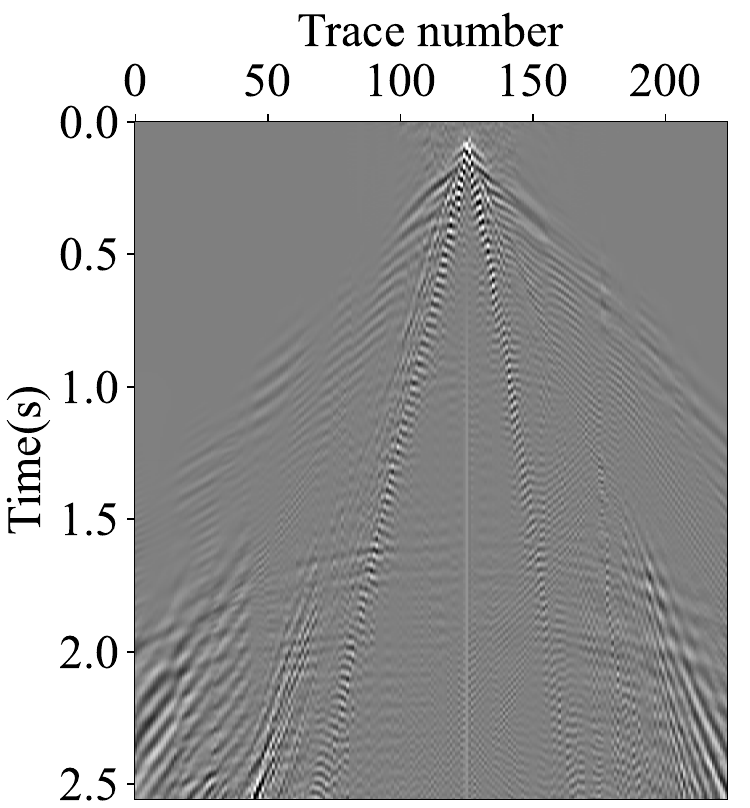}}
  \subfloat[\label{fig:arid_f}]{
		\includegraphics[width=0.5\columnwidth]{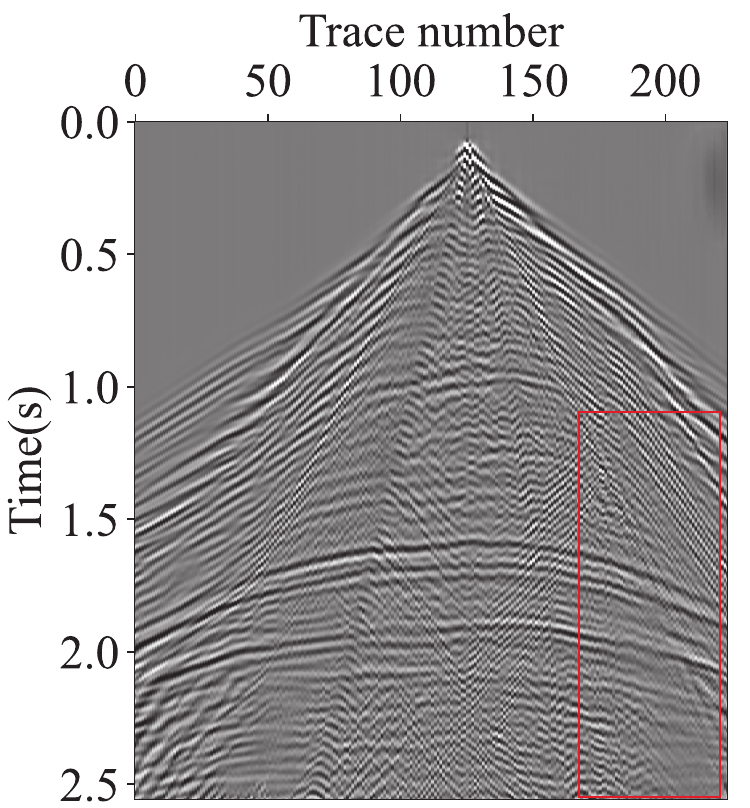}}
  \subfloat[\label{fig:arid_g}]{
		\includegraphics[width=0.5\columnwidth]{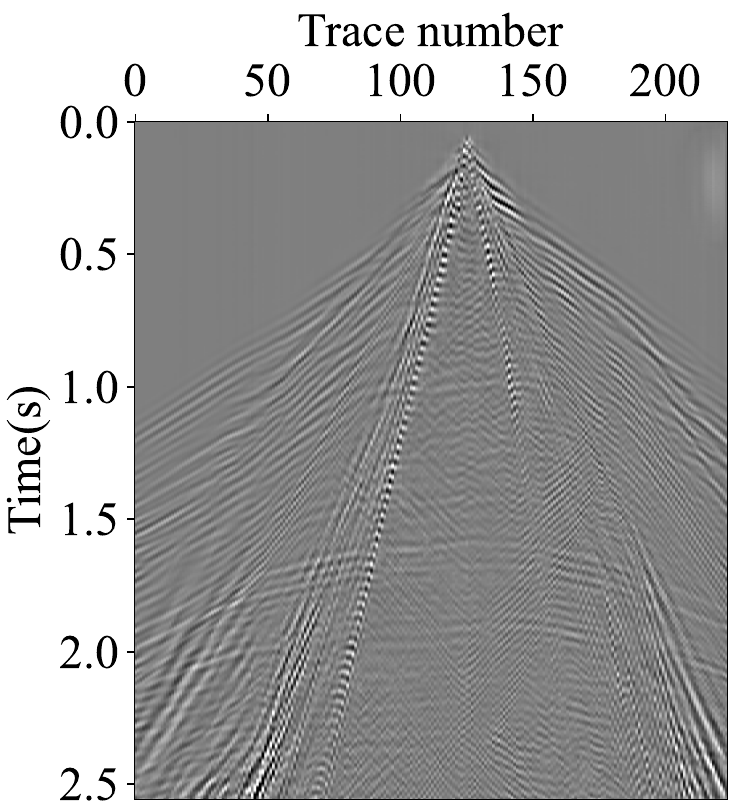}}\\
   \subfloat[\label{fig:arid_h}]{
		\includegraphics[width=0.5\columnwidth]{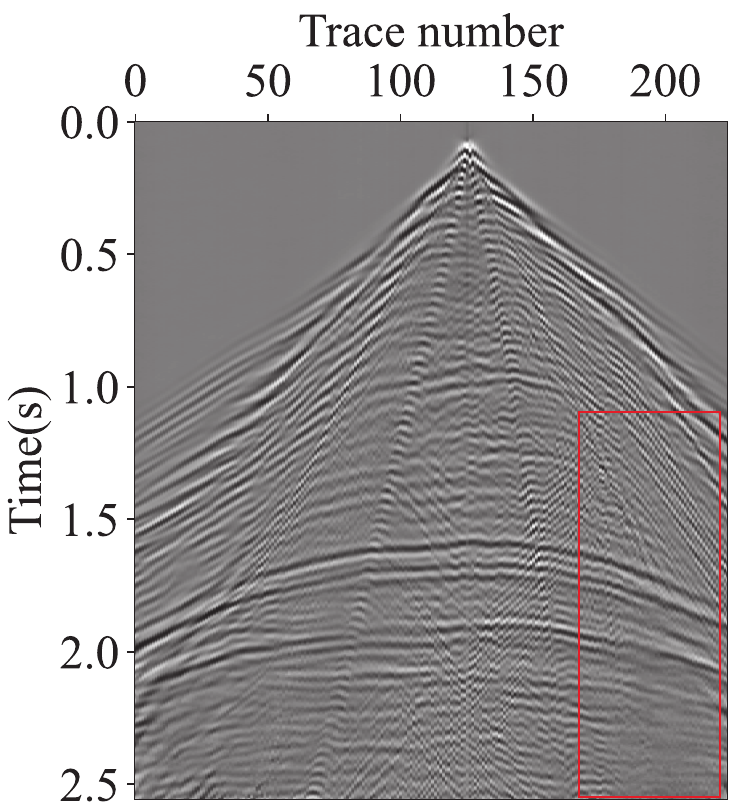}}	
   \subfloat[\label{fig:arid_i}]{
		\includegraphics[width=0.5\columnwidth]{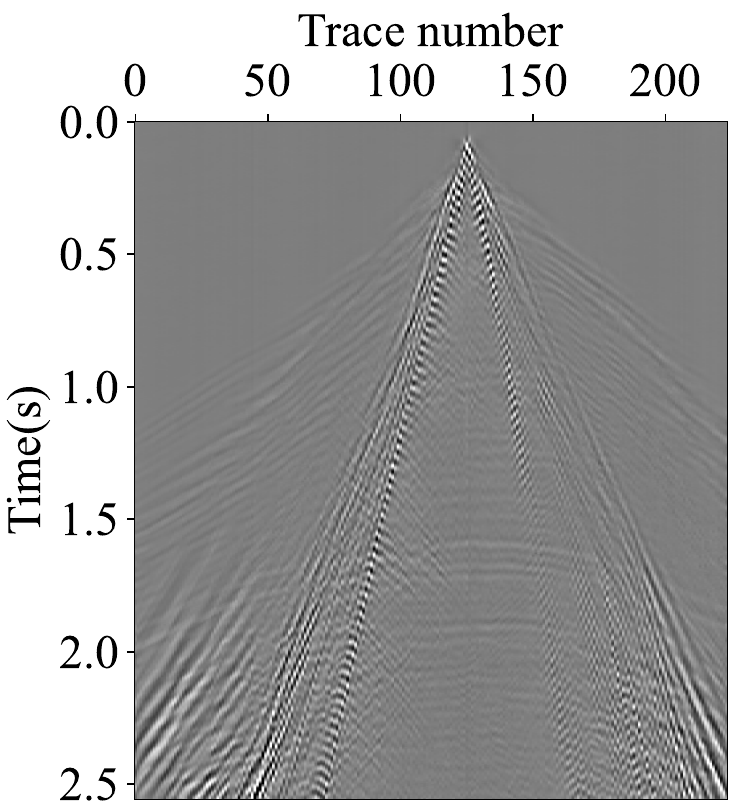}}
  \subfloat[\label{fig:arid_j}]{
		\includegraphics[width=0.5\columnwidth]{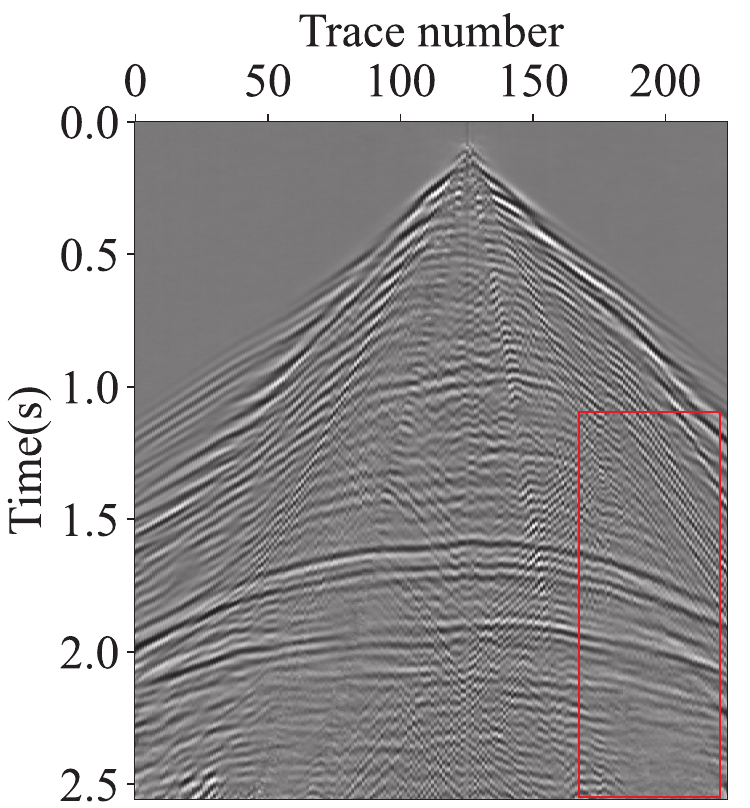}}
  \subfloat[\label{fig:arid_k}]{
		\includegraphics[width=0.5\columnwidth]{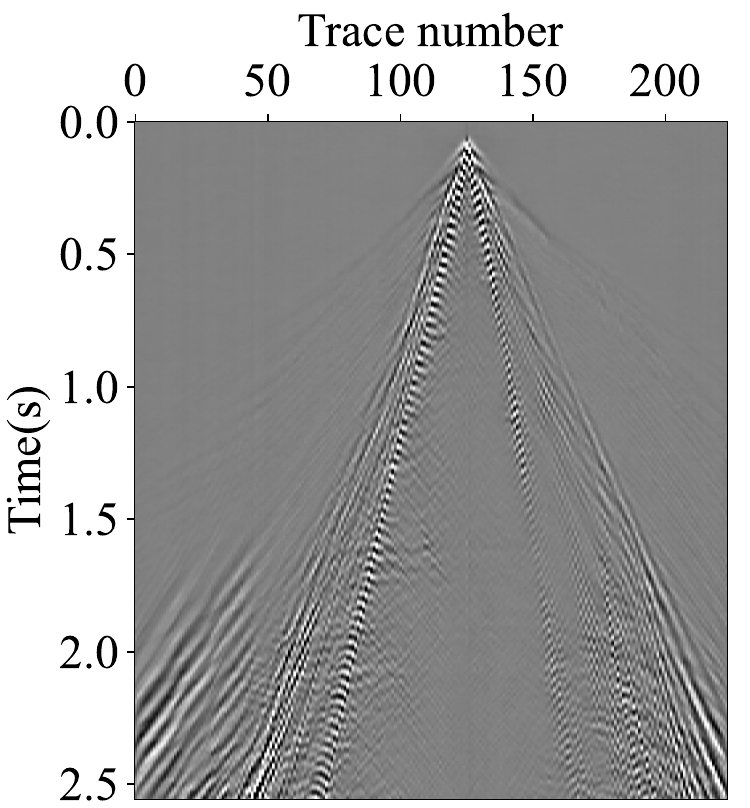}}\\
  \caption{Examples on the Arid synthetic data. (a) Noisy data. (b) True clean data. (c) True ground roll data. (d) Predicted clean data using LTF . (e) Predicted ground roll data using LTF. (f) Predicted clean data using U-Net. (g) Predicted ground roll data using U-Net. (h) Predicted clean data using DDPM-1c. (i) Predicted ground roll data using DDPM-1c. (j) Predicted clean data using DDPM-2c. (k) Predicted ground roll data using DDPM-2c.}
    \label{fig:arid}
\end{figure*}

\subsection{SEAM Arid synthetic example}
  We first test the proposed methods DDPM-1c and DDPM-2c on the synthetic data example. Besides, we also apply the LTF and U-Net to the synthetic data as comparison. Fig. \ref{fig:aridm} shows the 2D slice of the SEAM Arid model. To provide the benchmark for comparison, we can directly obtain the clean data and the ground roll by applying the elastic finite-difference modeling method to the modified velocity model. We use the \(v_p\) and \(\rho\) model shown in Fig. \ref{fig:aridm_a} and \ref{fig:aridm_c} but set \(v_s\) to zero to simulate the clean data, while we retain the near-surface part above 600 m of Fig. \ref{fig:aridm} but make the model beneath 600 m homogeneous to simulate the ground roll. The 25-Hz Ricker wavelet is used to simulate these seismic data. The simulated clean data and ground roll are shown in Fig. \ref{fig:arid_b} and Fig. \ref{fig:arid_c}, respectively. The simulated data has a total of 640 traces and 224 time samples per trace, with a time sampling interval of 0.004 s. We first apply automatic gain control (AGC) to these simulated data. Then, we add the weighted ground-roll data to the clean data to obtain the noisy data shown in Fig. \ref{fig:arid_a}. The weighting factor for the ground roll can be used to adjust the SNR of noisy data. We can see from Fig. \ref{fig:arid_a} that the coherent ground-roll noise corrupts the shallow reflections at near offsets and deeper reflections at far offsets. We use our proposed method (DDPM-1c, DDPM-2c), LTF and U-Net to remove the ground roll from the noisy data. Fig. \ref{fig:arid_d} and \ref{fig:arid_e} show the clean data and ground roll obtained by the LTF method, respectively. We can obviously see some ground-roll noise in the denoising result. The clean data and ground roll corresponding to the U-Net method are shown in Fig. \ref{fig:arid_f} and \ref{fig:arid_g}. 
  We can still see that the ground roll is not removed completely, especially for the part pointed by the red box. Besides, the energy leakage of reflections is observed in Fig. \ref{fig:arid_g}. The clean data predicted by using DDPM-1c is shown in Fig. \ref{fig:arid_h}. As pointed by the red box, the ground-roll energy leakage is relatively weak. The separated ground-roll data by using DDPM-1c is shown in Fig. \ref{fig:arid_i}. We can see that the energy leakage of reflections is less visible compared with LTF and U-Net. Fig. \ref{fig:arid_j} and \ref{fig:arid_k} show the predicted clean data and ground roll with DDPM-2c. We can see that the predicted clean data is comparable to the simulated clean with very slight contamination from ground roll. As pointed by the red box, the ground roll is completely removed, and the masked reflections are recovered well. In addition, there is obviously less remaining reflection energy in the predicted ground roll compared with the result of DDPM-1c. The comparison between these figures tell us that the proposed DDPM-2c shows better performance in ground-roll attenuation than LTF, U-Net and DDPM-1c.

\begin{figure} [htp!]
	\centering
 \subfloat[\label{fig:arid_spec2_a}]{
		\includegraphics[width=0.5\columnwidth]{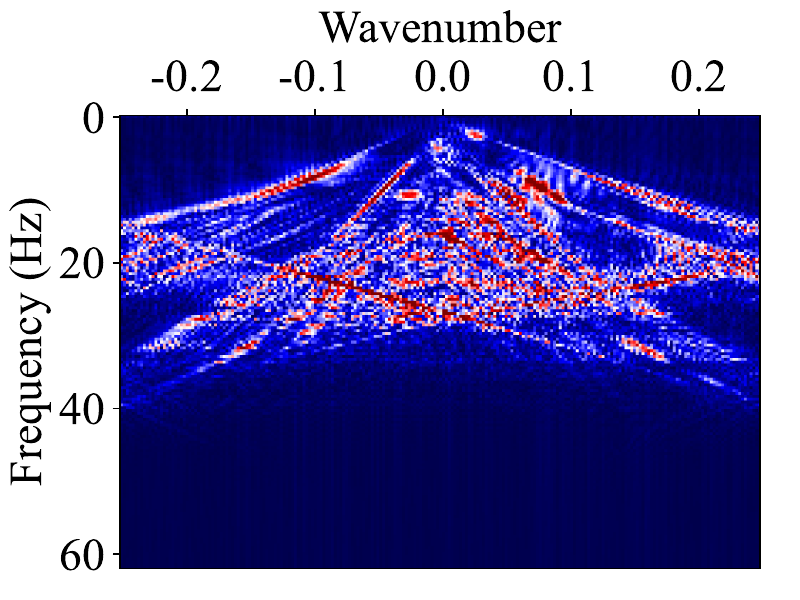}}	
   \subfloat[\label{fig:arid_spec2_b}]{
		\includegraphics[width=0.5\columnwidth]{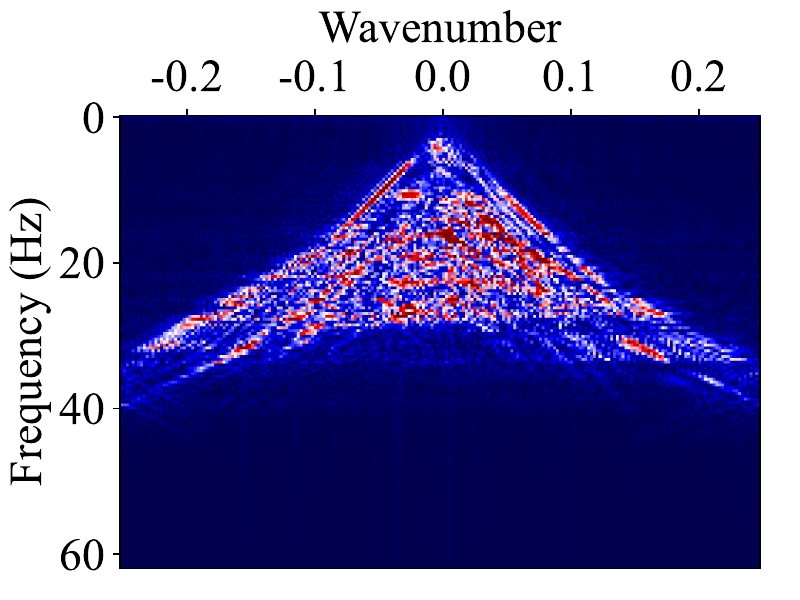}}\\
   \subfloat[\label{fig:arid_spec2_c}]{
		\includegraphics[width=0.5\columnwidth]{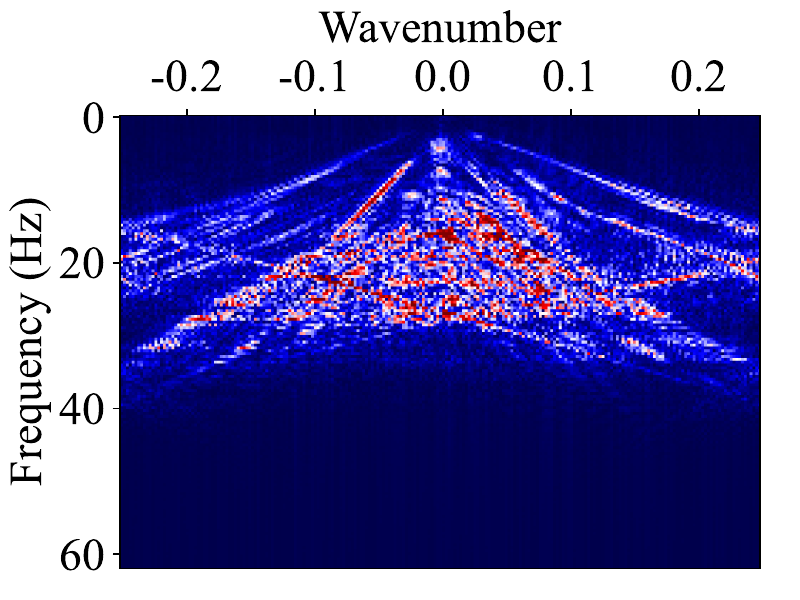}}	
  \subfloat[\label{fig:arid_spec2_d}]{
		\includegraphics[width=0.5\columnwidth]{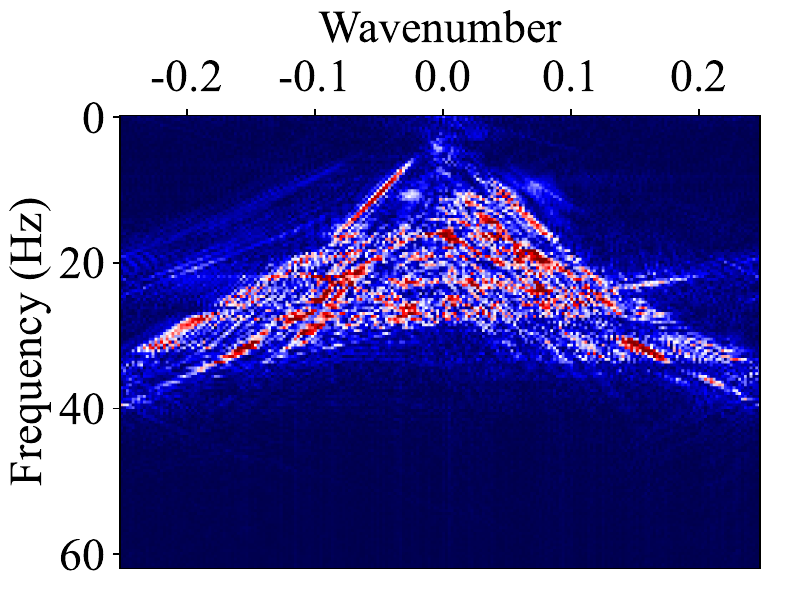}}\\
   \subfloat[\label{fig:arid_spec2_e}]{
		\includegraphics[width=0.5\columnwidth]{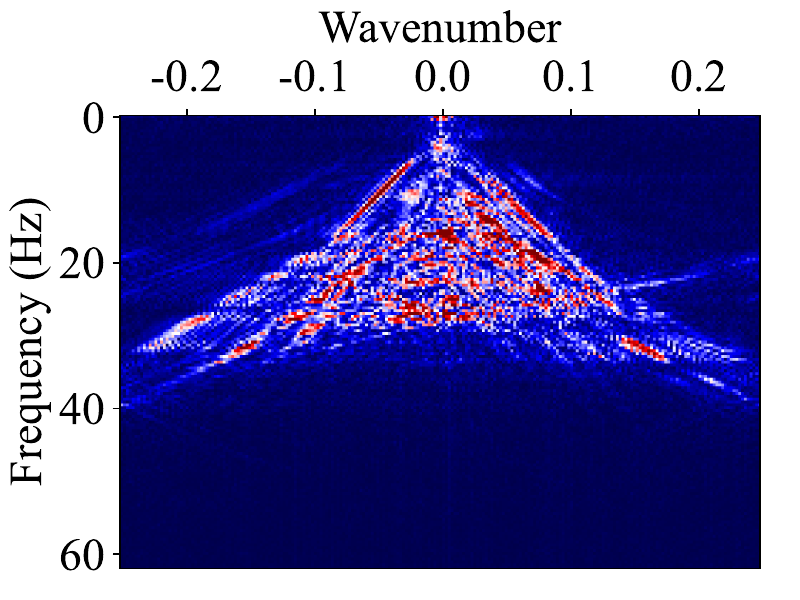}}
  \subfloat[\label{fig:arid_spec2_f}]{
		\includegraphics[width=0.5\columnwidth]{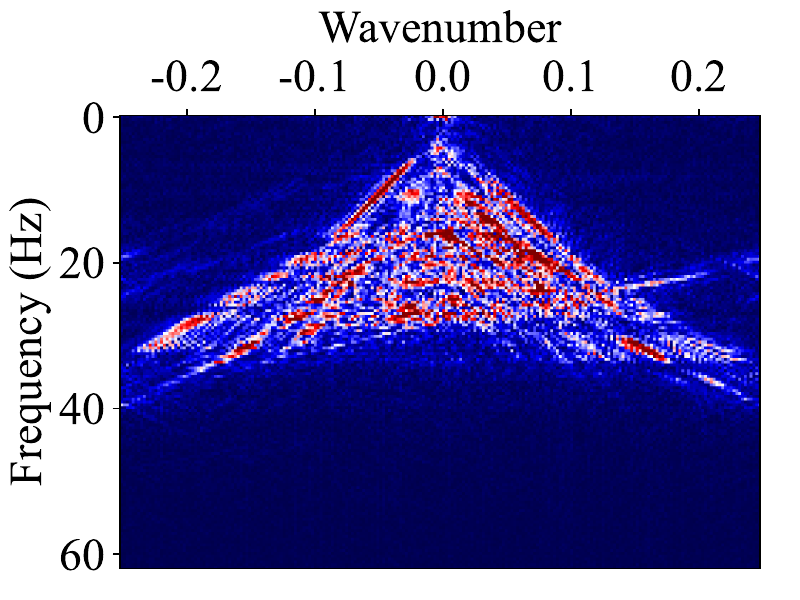}}\\
 \caption{F-K planes of the Arid synthetic data in Fig. \ref{fig:arid}. (a) Noisy data. (b) True clean data. (c) Predicted clean data using LTF. (d) Predicted clean data using U-Net. (e) Predicted clean data using DDPM-1c. (f) Predicted clean data using DDPM-2c. }
    \label{fig:spec2_arid}
\end{figure}

\begin{figure} [htp!]
	\centering
  \subfloat[\label{fig:spec1_arid_a}]{
		\includegraphics[width=1\columnwidth]{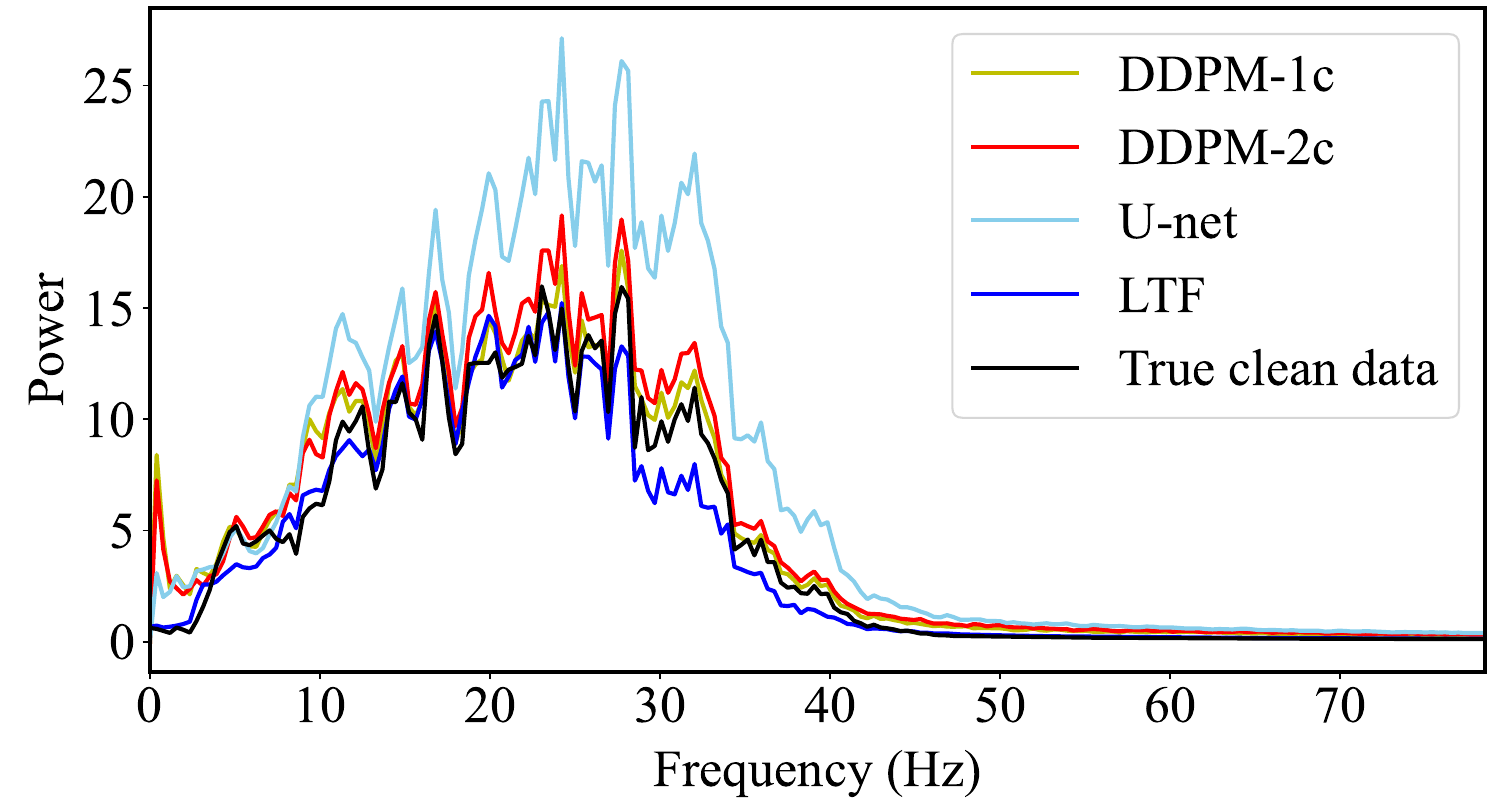}}\\
  \subfloat[\label{fig:spec1_arid_b}]{
		\includegraphics[width=1\columnwidth]{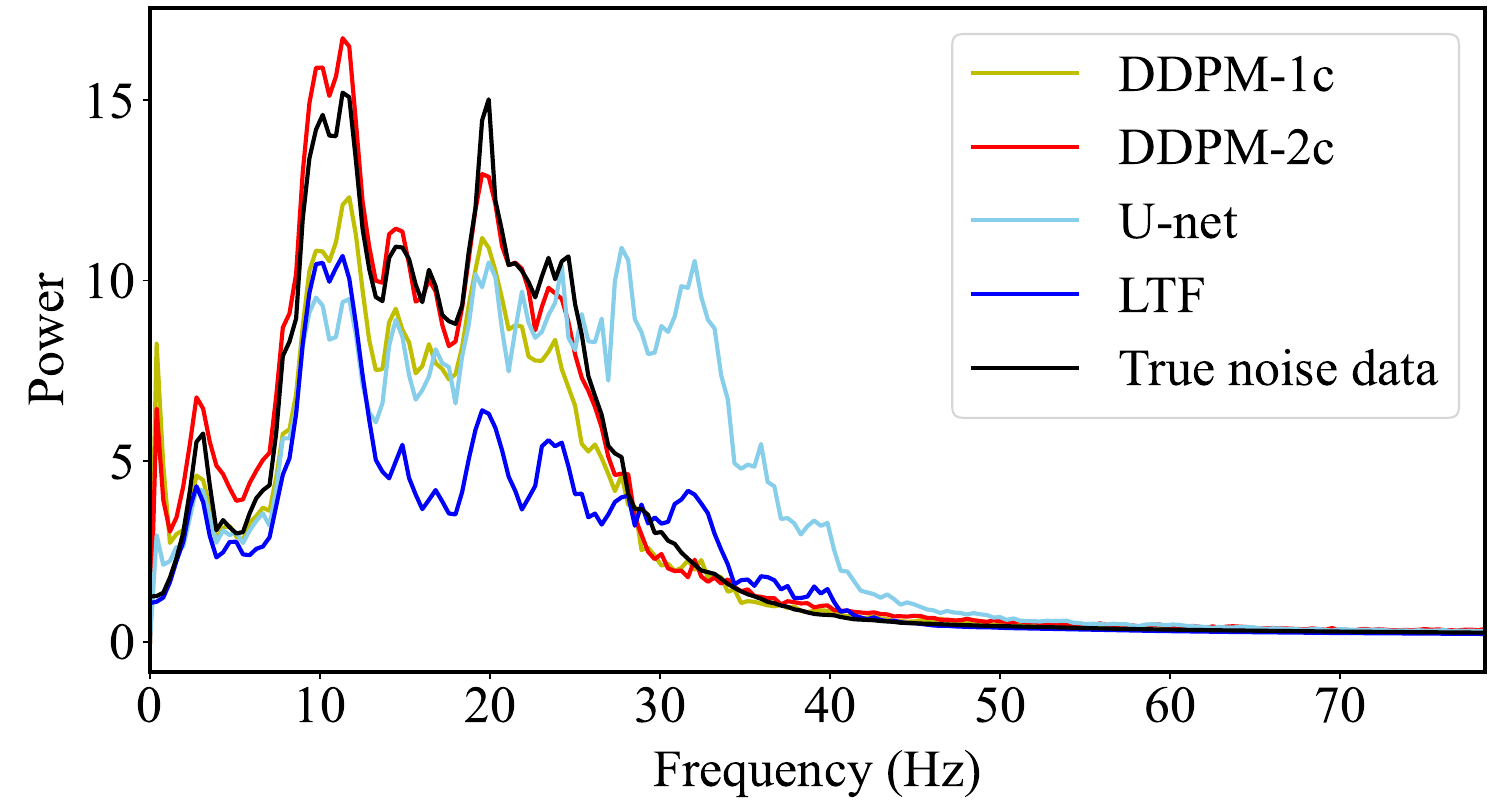}} 
  \caption{Amplitude spectra comparison of the Arid synthetic data example in Fig. \ref{fig:arid}. (a) The clean data. (b) The ground roll data. }
  \label{fig:spec1_arid}
\end{figure}

 We apply the F-K plane analysis to further evaluate the performance of the ground-roll attenuation methods. The F-K spectra corresponding to the data in Fig. \ref{fig:arid_a} and Fig. \ref{fig:arid_b} are shown in Fig. \ref{fig:arid_spec2_a} and \ref{fig:arid_spec2_b}, respectively. From the spectra, we observe that the clean data and ground roll have overlap in the frequency band of 20 - 35Hz. Fig. \ref{fig:arid_spec2_c} displays the F-K spectra of the predicted clean data in Fig. \ref{fig:arid_d}, containing some residual ground-roll energy. Fig. \ref{fig:arid_spec2_d}, \ref{fig:arid_spec2_e}, and \ref{fig:arid_spec2_f} show the F-K spectra corresponding to the predicted clean data in Fig. \ref{fig:arid_f}, \ref{fig:arid_h}, and \ref{fig:arid_j}, respectively. The F-K spectra corresponding to DDPM-2c has less residual ground roll and looks closer to that of the simulated clean data.

Moreover, we utilize amplitude spectra to assess the performance of these methods for attenuating ground roll in noisy data. The amplitude spectra of the true clean data and the predicted clean data in Fig. \ref{fig:arid} are illustrated in Fig. \ref{fig:spec1_arid_a}. We compute the spectra of these data over the same time and space window. Obviously, the U-Net result has an increased energy from 10 to 40 Hz, arising from the inaccurate amplitude of the predicted clean data. On the other hand, the clean data predicted by LTF, DDPM-1c and DDPM-2c have consistent frequencies with the true clean data. Furthermore, the amplitude spectra of the true noise data and the separated ground-roll data in Fig. \ref{fig:arid} are shown in Fig. \ref{fig:spec1_arid_b}. We can see that amplitude spectra corresponding to DDPM-2c shows the best consistency with the target spectra. Although the prediction of DDPM-1c slightly degrades, it is still better than the LTF and U-Net methods.  
 \begin{table}
\caption{Evaluations for the predicted clean data}
\begin{tabular}{cccccc}
   \toprule
   Methods & MAE &  MSE & SSIM & PSNR  \\
   \midrule
   LTF & 0.0229 & 0.0014 & 0.7463 & 28.4057 \\
   U-Net & 0.0328 & 0.0026 & 0.7272 & 25.7754 \\
   DDPM-1c & 0.0189 & 0.0009 & 0.7426 & 30.3792\\
   DDPM-2c & 0.0191 & 0.0009 & 07807 & 30.1405\\

   \bottomrule
\end{tabular}
\label{table:hr}
\end{table}

\begin{table}
\caption{Evaluations for the predicted noise data}

\begin{tabular}{cccccc}
   \toprule
   Methods & MAE & MSE & SSIM & PSNR  \\
   \midrule
   LTF & 0.0229 & 0.0014 & 0.5991 & 28.4307 \\
   U-Net & 0.0328 & 0.0026 & 0.4516 & 25.7766\\
   DDPM-1c & 0.0188 & 0.0009 & 0.6084 & 30.4044\\
   DDPM-2c & 0.0174 & 0.0008 & 0.7038 & 30.5280\\

   \bottomrule
\end{tabular}
\label{table:lr}
\end{table}

For a quantitation assessment, we then use four evaluation metrics, including Mean absolute error (MAE), mean square error (MSE), peak signal-to-noise ratio (PSNR) and structural similarity index measure (SSIM) to measure the difference between the ground truth and predicted results. You are referred to Appendix A for more details about these evaluation metrics. The lower the values of MAE and MSE, and the higher the values of PSNR and SSIM, the closer the predicted data is to the target. Considering that the DDPMs has the inherent generative diversity, we perform the sampling process for the DDPM-based methods five times and compute the mean value of the metrics. The evaluation results for the clean data and ground roll predicted by different methods are summarized in Table. \ref{table:hr} and \ref{table:lr}, respectively. We can see from Table. \ref{table:hr} that our proposed methods (DDPM-1c and DDPM-2c) perform better than the LTF and U-Net methods in predicting clean data. Meanwhile, Table. \ref{table:lr} illustrates that DDPM-2c has superior performance in predicting ground roll. 

\subsection{Field data example}
\begin{figure*} [htp!]
	\centering
 \subfloat[\label{fig:field_a}]{
		\includegraphics[width=0.5\columnwidth]{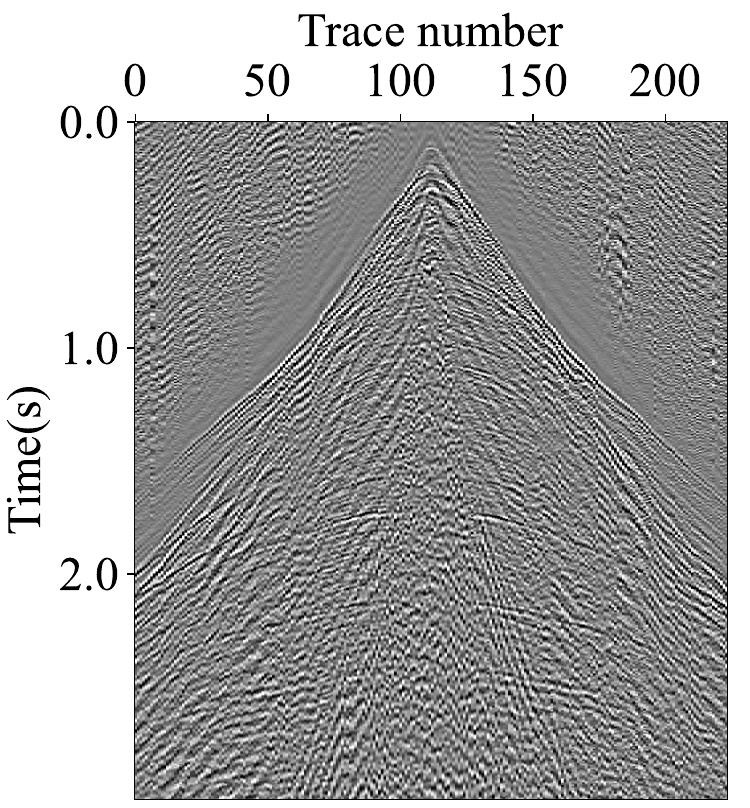}}\\
   \subfloat[\label{fig:field_b}]{
		\includegraphics[width=0.5\columnwidth]{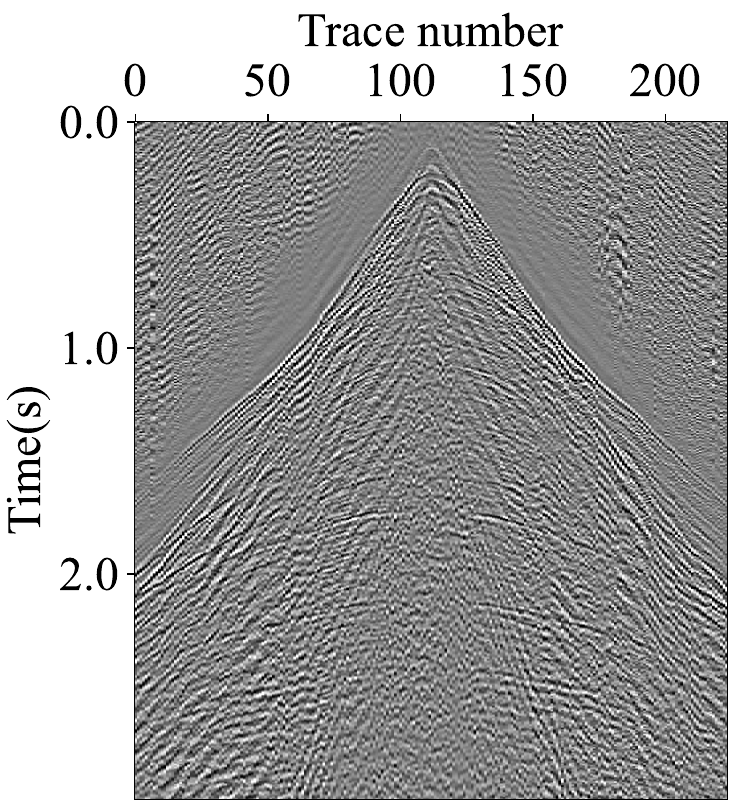}}	
   \subfloat[\label{fig:field_c}]{
		\includegraphics[width=0.5\columnwidth]{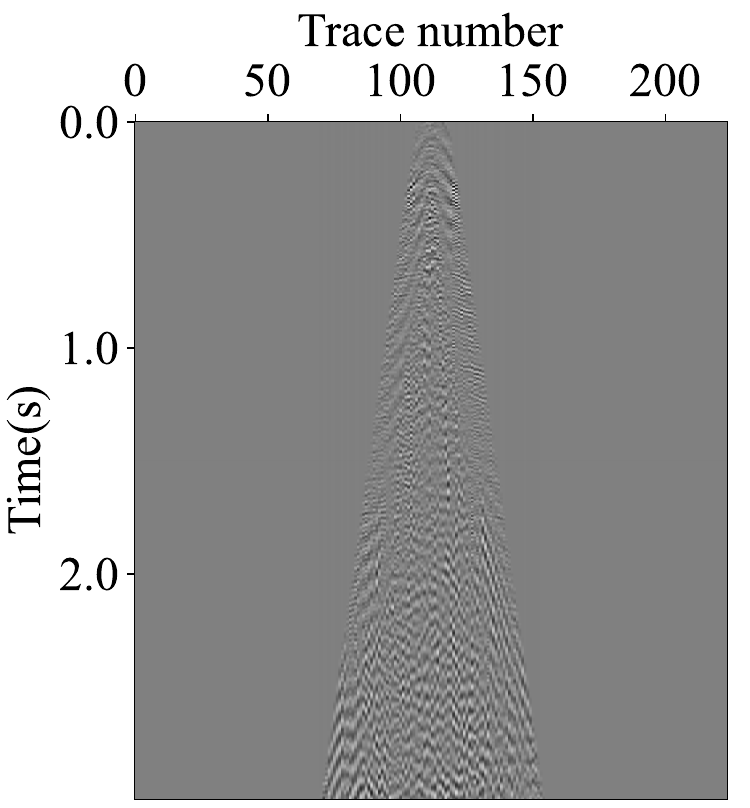}}
  \subfloat[\label{fig:field_d}]{
		\includegraphics[width=0.5\columnwidth]{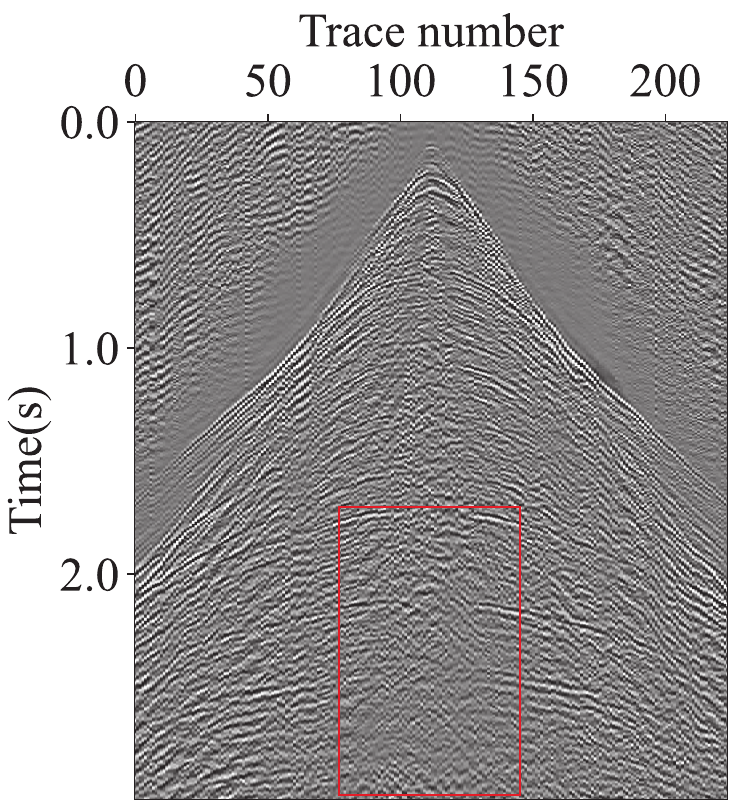}}
  \subfloat[\label{fig:field_e}]{
		\includegraphics[width=0.5\columnwidth]{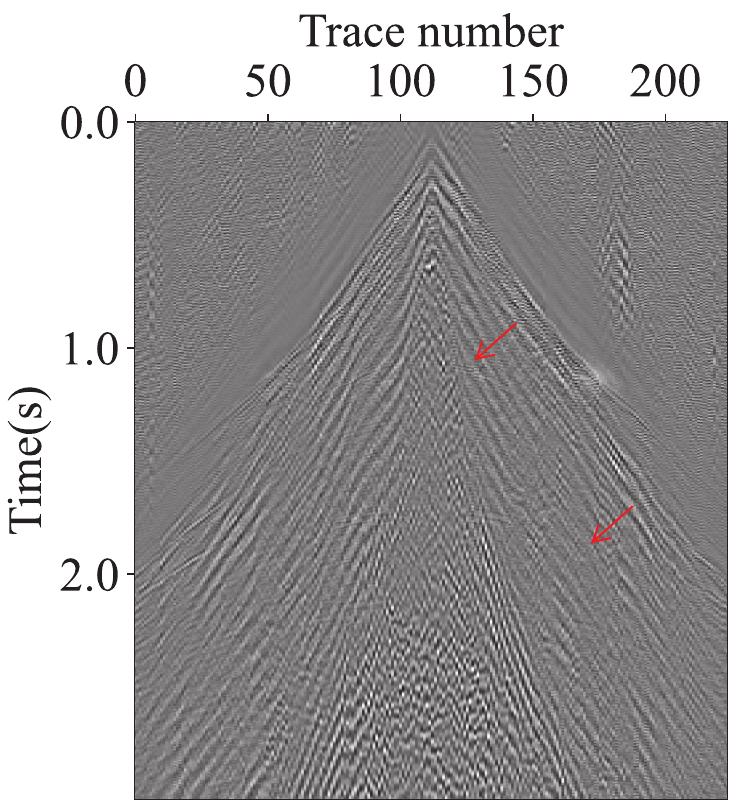}}\\
   \subfloat[\label{fig:field_f}]{
		\includegraphics[width=0.5\columnwidth]{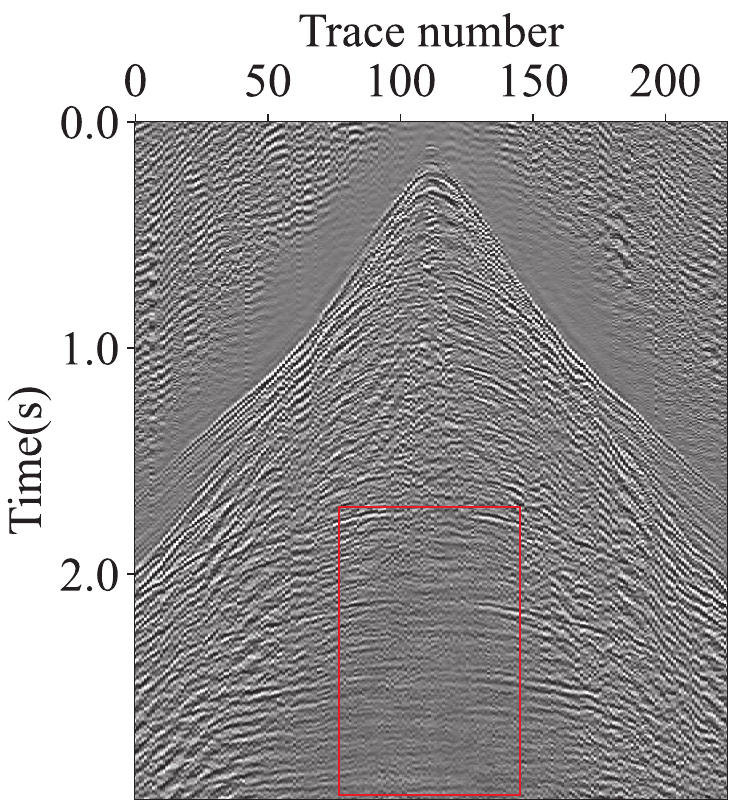}}	
   \subfloat[\label{fig:field_g}]{
		\includegraphics[width=0.5\columnwidth]{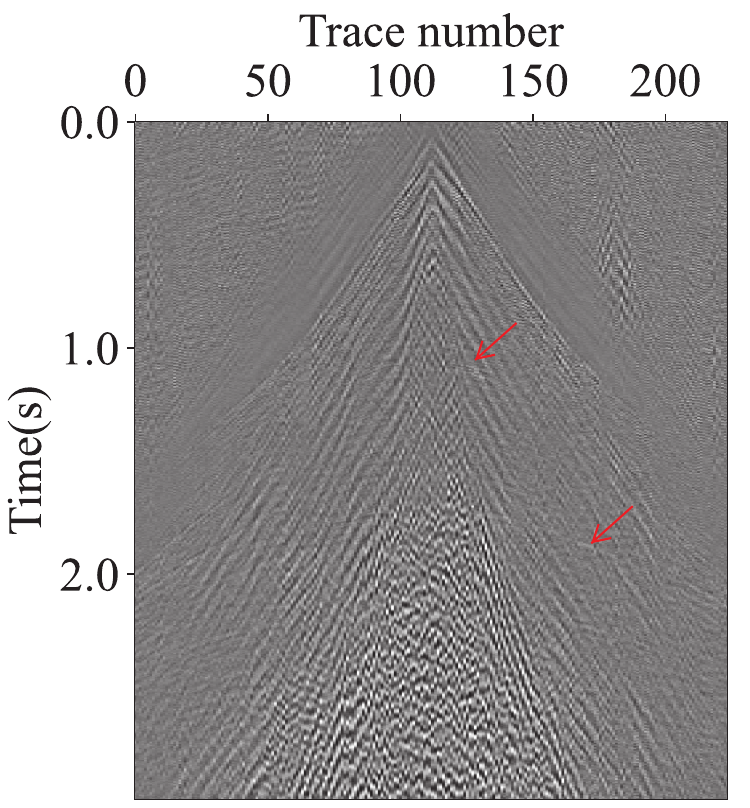}}
  \subfloat[\label{fig:field_h}]{
		\includegraphics[width=0.5\columnwidth]{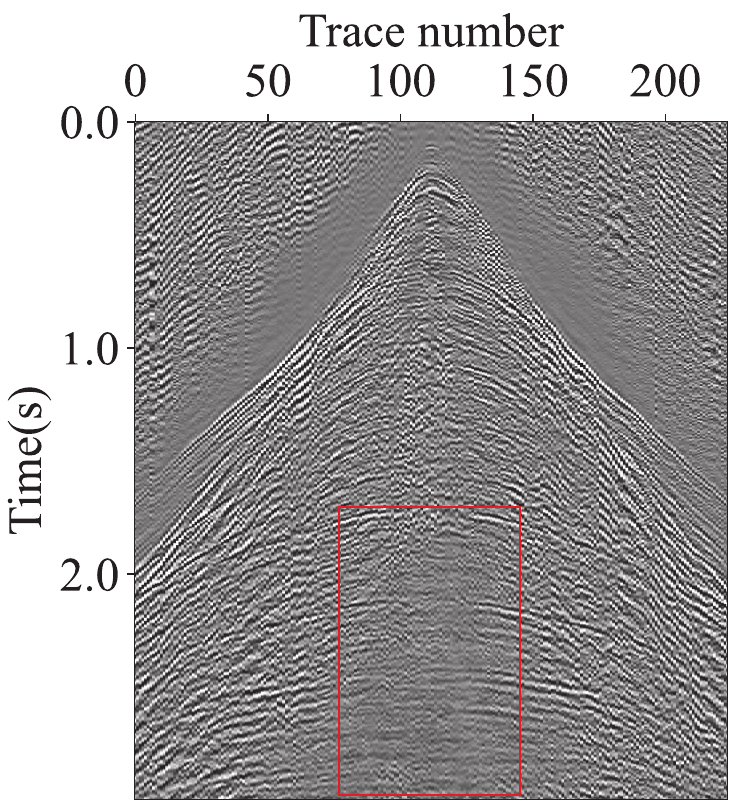}}
  \subfloat[\label{fig:field_i}]{
		\includegraphics[width=0.5\columnwidth]{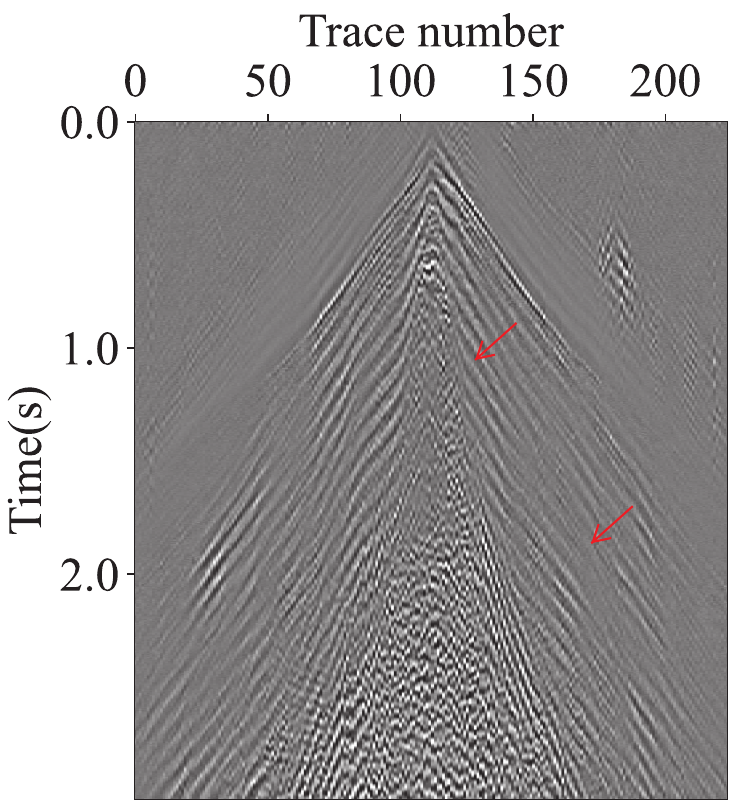}}\\
  \caption{Examples on the field data. (a) Noisy data.  (b) Predicted clean data using LTF. (c) Predicted ground roll data using LTF. (d) Predicted clean data using U-Net. (e) Predicted ground roll data using U-Net. (f) Predicted clean data using DDPM-1c. (g) Predicted ground roll data using DDPM-1c. (h) Predicted clean data using DDPM-2c. (i) Predicted ground roll data using DDPM-2c.}
    \label{fig:field}
\end{figure*}

We then apply the c-DDPM method to the field data with no needs to re-train the model. Fig. \ref{fig:field_a} shows a shot gather from the Line 001 dataset after AGC and down-sampling. The trace number is 640, and the recording length is 224 with a time sampling of 0.0047 s. The ground-roll noise is predominantly located at near offset and seriously contaminates the reflection events. The denoising result of the LTF method is shown in Fig. \ref{fig:field_b}. We can see that the reflections are still masked by strong ground-roll residuals. The ground roll corresponding to the LTF method is shown in Fig. \ref{fig:field_c}. There is obvious leakage of reflections at the masked area of the ground roll and reflections. The predicted clean data using U-Net is shown in Fig. \ref{fig:field_d}. As shown by the red box, the reflections are partially recovered, but the ground roll still obviously contaminates the seismic data. Fig. \ref{fig:field_e} shows the corresponding ground roll. We can see the ground roll is more complete, while there are some leakage of reflections as pointed by the red arrows. The clean data predicted by DDPM-1c and DDPM-2c are shown in Fig. \ref{fig:field_f} and Fig. \ref{fig:field_h}, respectively. As pointed by the red box, the reflections are recovered better than the LTF and U-Net methods. The predicted ground roll by using DDPM-1c and DDPM-2c are shown in Fig. \ref{fig:field_g} and Fig. \ref{fig:field_i}, respectively. As pointed out by the red arrows, we can see the slight reflection energy in the result of DDPM-1c, while the result of DDPM-2c has no visible reflections. These figures show that the DDPM-2c has better performance on the field data than the LTF, U-Net and DDPM-1c methods.

\begin{figure} [htp!]
	\centering
 \subfloat[\label{fig:spec2_line_a}]{
		\includegraphics[width=0.5\columnwidth]{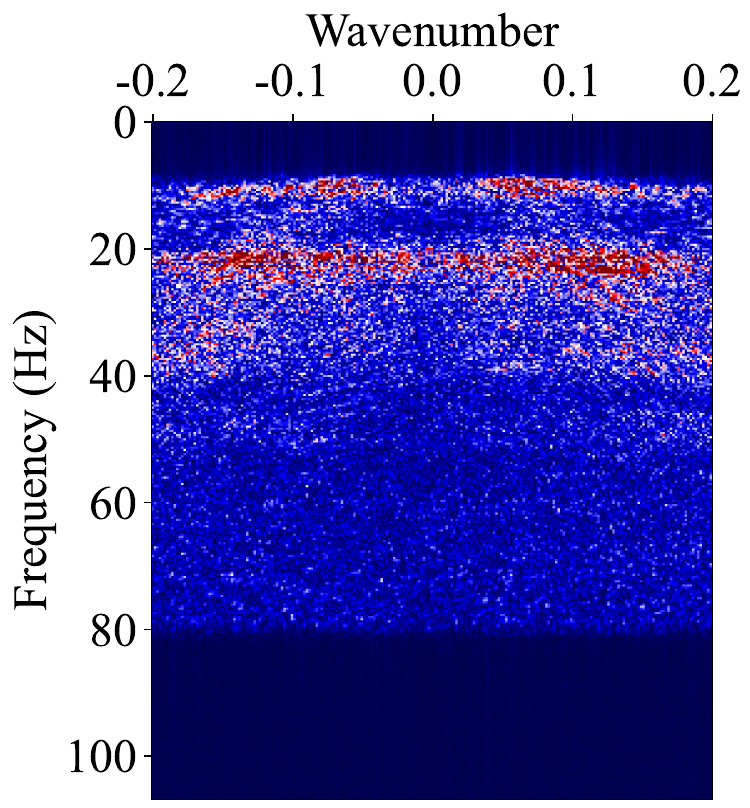}}\\
   \subfloat[\label{fig:spec2_line_b}]{
		\includegraphics[width=0.5\columnwidth]{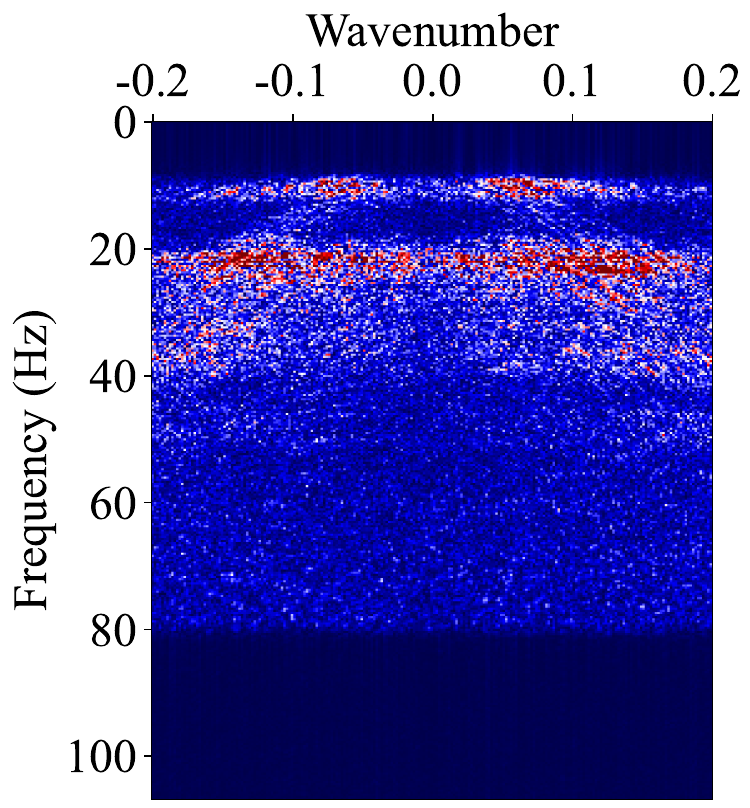}}	
  \subfloat[\label{fig:spec2_line_c}]{
		\includegraphics[width=0.5\columnwidth]{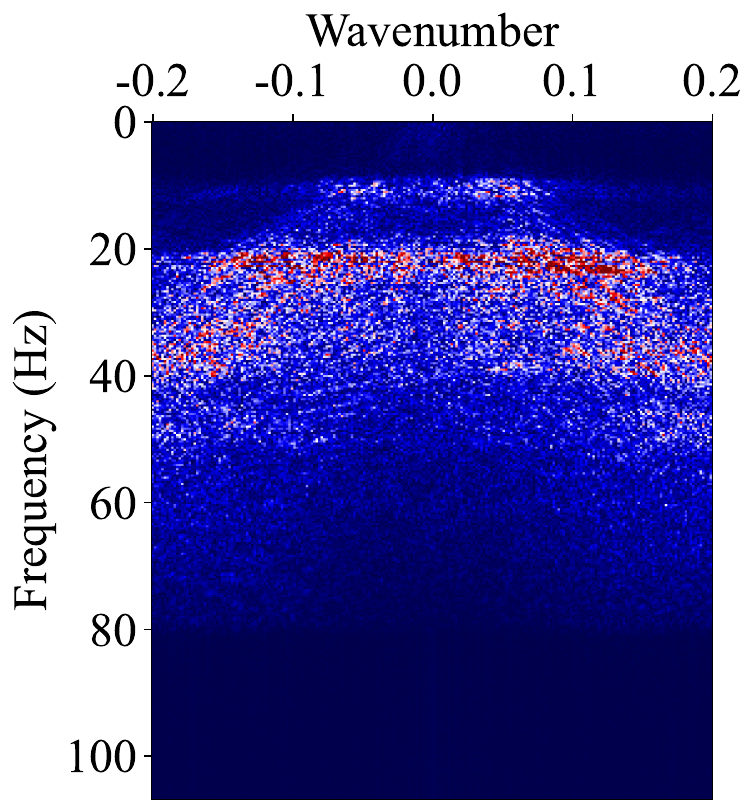}}\\
   \subfloat[\label{fig:spec2_line_d}]{
		\includegraphics[width=0.5\columnwidth]{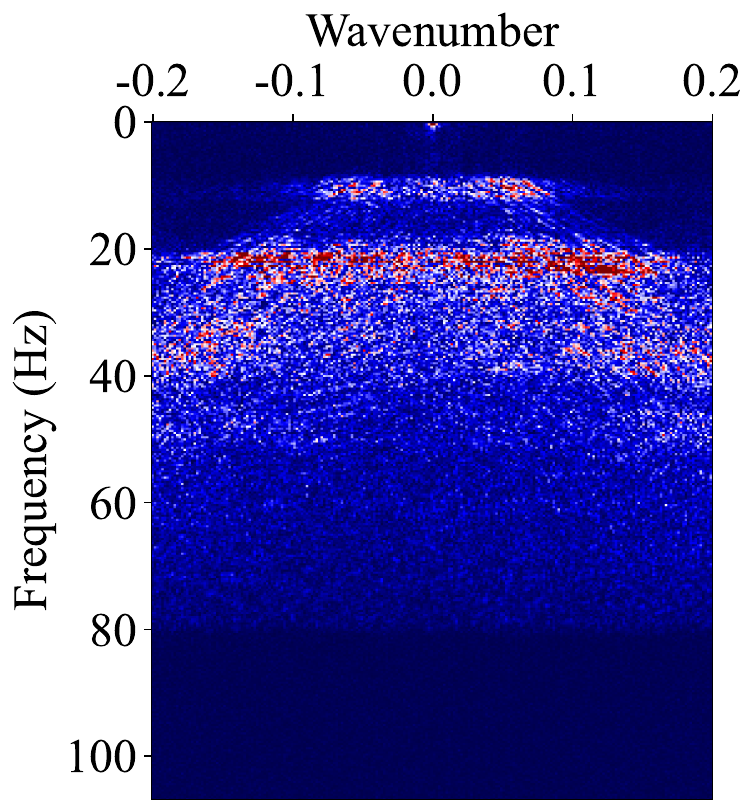}}	
  \subfloat[\label{fig:spec2_line_e}]{
		\includegraphics[width=0.5\columnwidth]{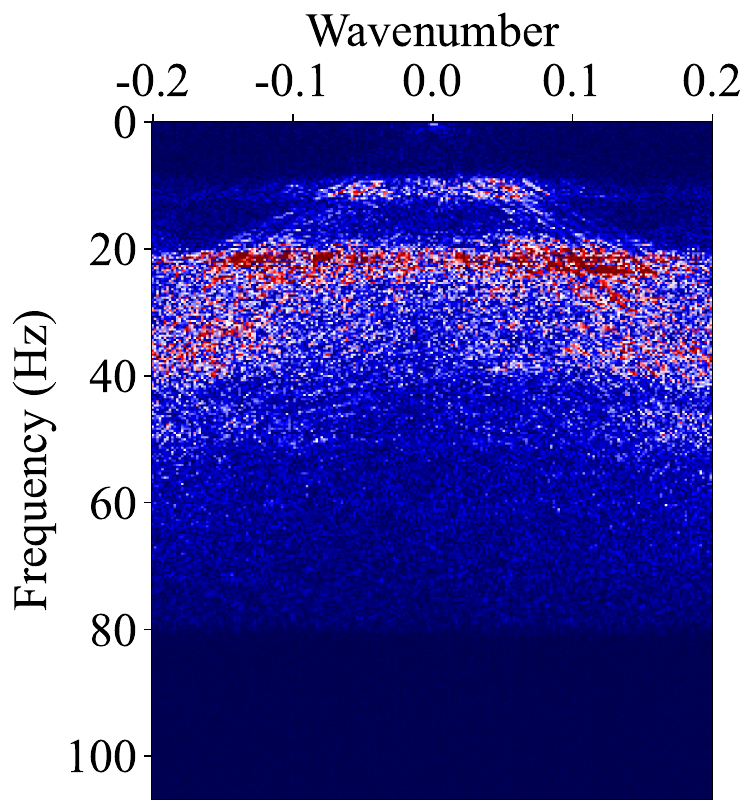}}\\
  \caption{F-K plane of the field data example. (a) Noisy data. (b) Predicted clean data using LTF. (c) Predicted clean data using U-Net. (d) Predicted clean data using DDPM-1c. (e) Predicted clean data using DDPM-2c.}
    \label{fig:spec2_line}
\end{figure}

\begin{figure} [htp!]
	\centering
  \subfloat[\label{fig:spec1_line_a}]{
		\includegraphics[width=1\columnwidth]{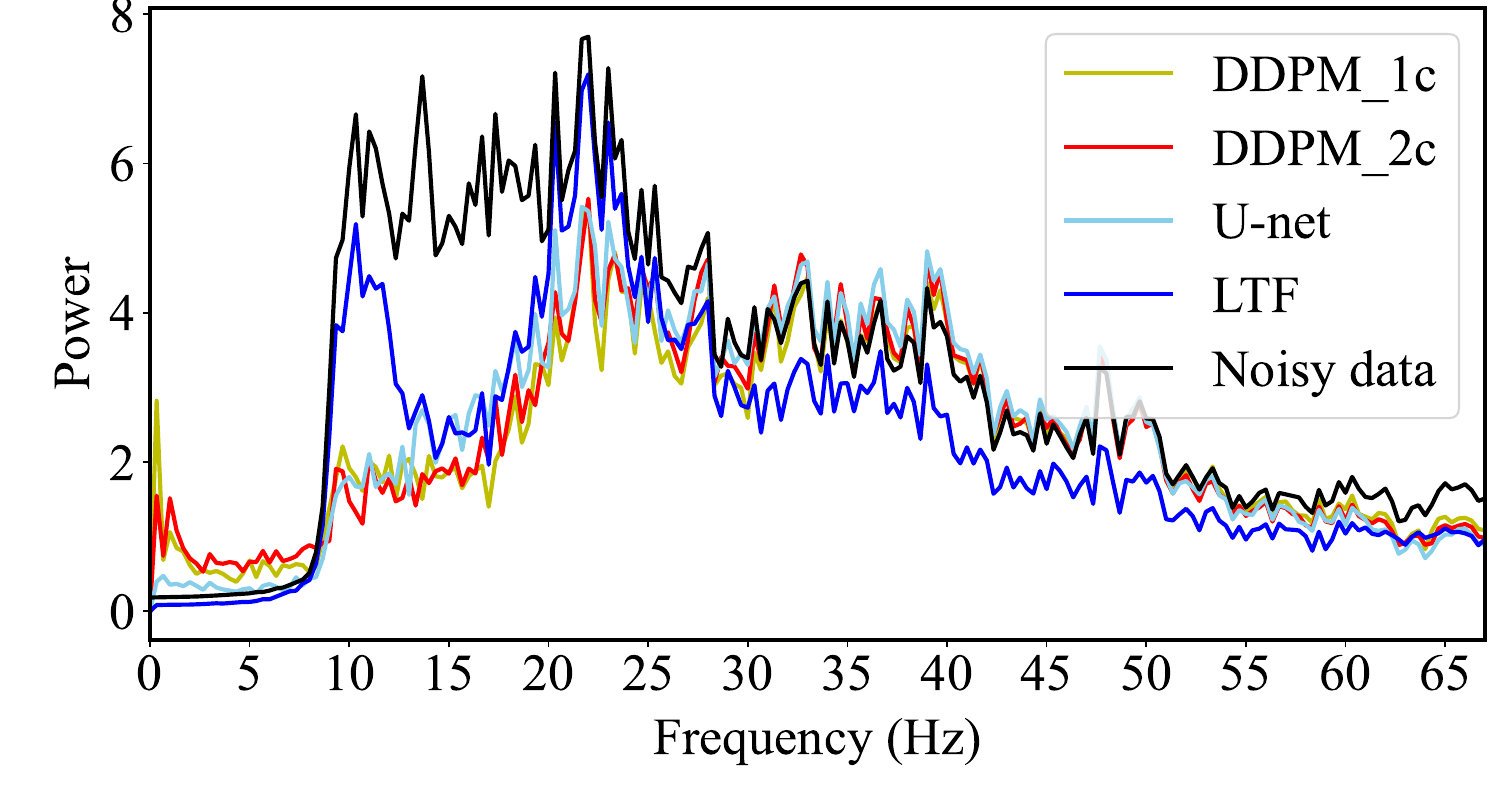}}\\
  \subfloat[\label{fig:spec1_line_b}]{
		\includegraphics[width=1\columnwidth]{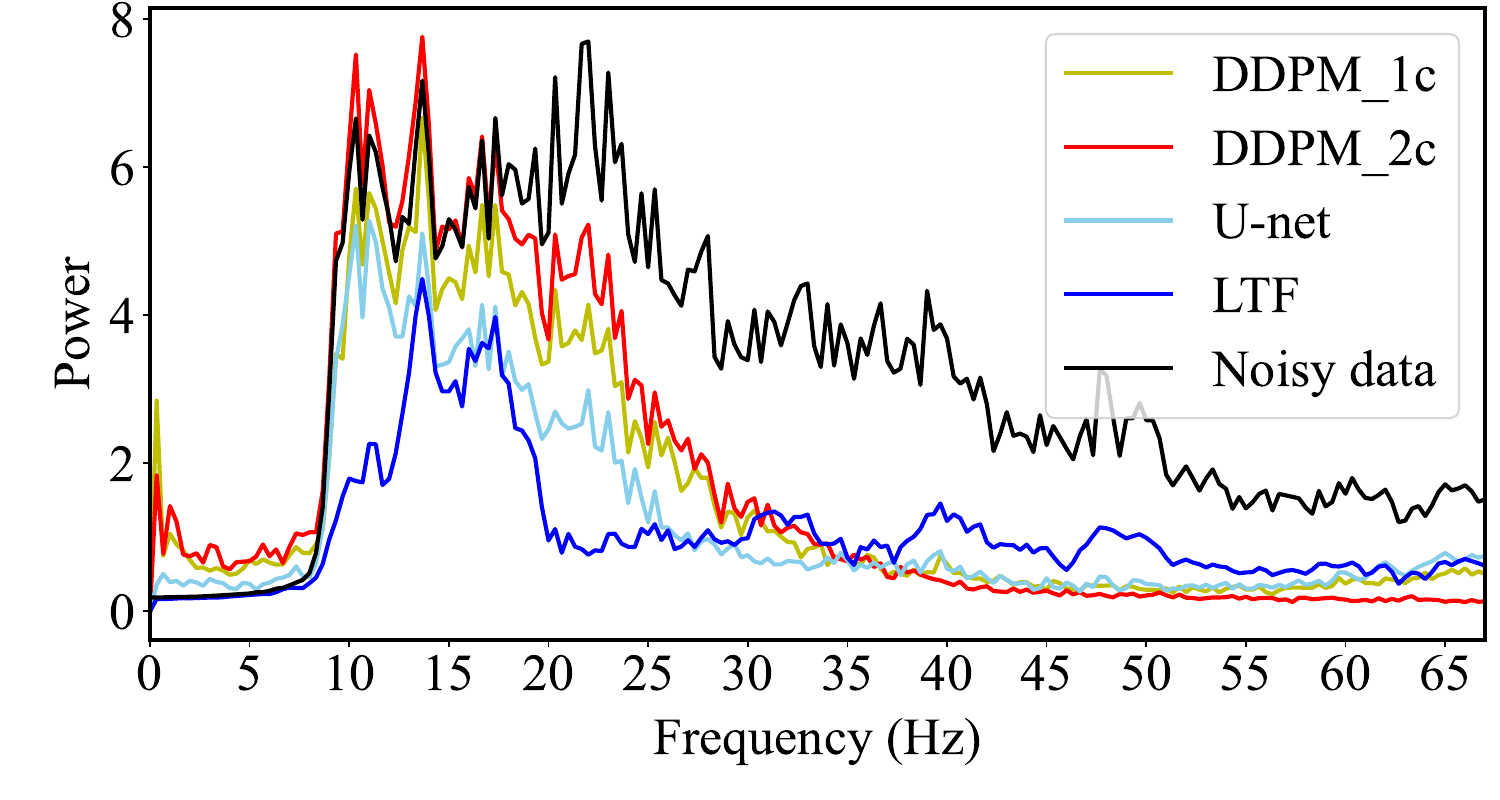}}\\
  \caption{Amplitude spectra comparison of the field data example
in Fig. 4. (a) The clean data. (b) The ground roll data. Black, red, yellow, sky-blue, blue lines indicate the spectra of noisy input data, separated clean data using DDPM-2c, DDPM-1c, U-Net and LTF, respectively.}
    \label{fig:spec1_line}
\end{figure}

We also show the f-k spectra of the original seismic data and the denoising results of different methods in Fig. \ref{fig:spec2_line}. The amplitude spectra of the original seismic data and denoising results are shown in Fig. \ref{fig:spec1_line_a}. We can see that the LTF does not attenuate the ground roll with frequency from 8 Hz to 15 Hz effectively. The amplitude spectra of the original seismic data and the predicted ground roll in Fig. \ref{fig:field} are shown in Fig. \ref{fig:spec1_line_b}. We can see that DDPM-1c and DDPM-2c predicts the ground roll with frequency from 10 to 20 Hz very well.
\setlength{\tabcolsep}{8pt}

\section{Discussion}

The proposed method (DDPM-1c and DDPM-2c) shows better performance on the synthetic and field data examples in ground-roll attenuation than the LTF and U-Net methods. Moreover, DDPM-2c can output the clean data and ground roll simultaneously with high quality. To make a further investigation of the proposed method, we then analyse the influence of AGC and training dataset on the denoising performance, along with the acceleration in diffusion denoising implicit model(DDIM).
\subsection{The influence of AGC}
It is more challenging when the energy of ground roll is much stronger than the reflections in the seismic data. Thus, we perform AGC on the noisy input in the synthetic and field data examples, to balance the amplitudes across the whole data. AGC adjust the amplitudes of seismic data in a window according to the mean energy within the window. The reflections overlapped by strong ground-roll noise will have smaller gain than the reflections unaffected by the ground-roll noise. Thus, the predicted reflections in the strongly-affected area tends to be weaker than the predicted reflections in the unaffected area. This can be seen in Fig. \ref{fig:field_f} and \ref{fig:field_h}.

There are two possible solutions to this issue:

1) Improve the network so that it can work well when the ground roll energy is dominant in the noisy data. DDPMs are deep generative models originally made for image synthesis. The network architecture for a range of 0-255 RGB of an image should be carefully modified for the high-resolution task such as seismic processing.

2) Reconstruct the clean data instead of separating it from the noisy data after AGC. Both the DDPM-1c and DDPM-2c are required to separate the reflection energy from the noisy condition after AGC. The network can learn to reconstruct the contaminated area with continuous reflection energy by lessening the constraint of the condition. In other words, the c-DDPM should possess more diversity to construct the contaminated area.
\subsection{The influence of training datasets}
The preparation of training dataset is an important component of the DL method. We applied the convolution modeling to generate the ground roll. We configure the source wavelet frequency for the convolution modeling to fall within a certain range, ensuring the trained model generalizes effectively across both synthetic and field data examples. The training dataset used above is named as "dataset1". To test how the frequency of the ground roll influence the generation quality,  we enlarge the frequency range of the ground roll and generate a new training dataset, named as "dataset2". The new dataset is then used to train the DDPM-2c with the same training parameters as before. After training, we then apply the trained model to the noisy field data in Fig. \ref{fig:field_a}. We can see that the residual ground-roll noise with higher frequency are also removed, but the direct arrivals are also missing. Additionally, there are more leakage of reflections into the predicted ground roll when using dataset2 for training compared with the scenario for dataset1. This is because the model trained on dataset2 tends to generate the events with higher frequency (which usually refer to the reflections), especially in the area where the reflections and the ground roll heavily overlap.

We then compute the amplitude spectra of the predicted clean data and ground roll corresponding to dataset2 to further demonstrate the influence of the frequency band of the ground roll for training on the data generation. The amplitude spectra of the predicted clean data using dataset2 exhibits an obvious decrease after 20Hz, while the amplitude spectra of the ground roll increases after 20Hz. The experiment result demonstrates that the frequency band of the training dataset affects the generation quality of the ground roll and the clean data. Thus, we should carefully prepare the training dataset with a reasonable frequency band to improve the predicted clean data and ground roll. 
\begin{figure} [htp!]
	\centering
 \subfloat[\label{fig:lowf_line_a}]{
		\includegraphics[width=0.5\columnwidth]{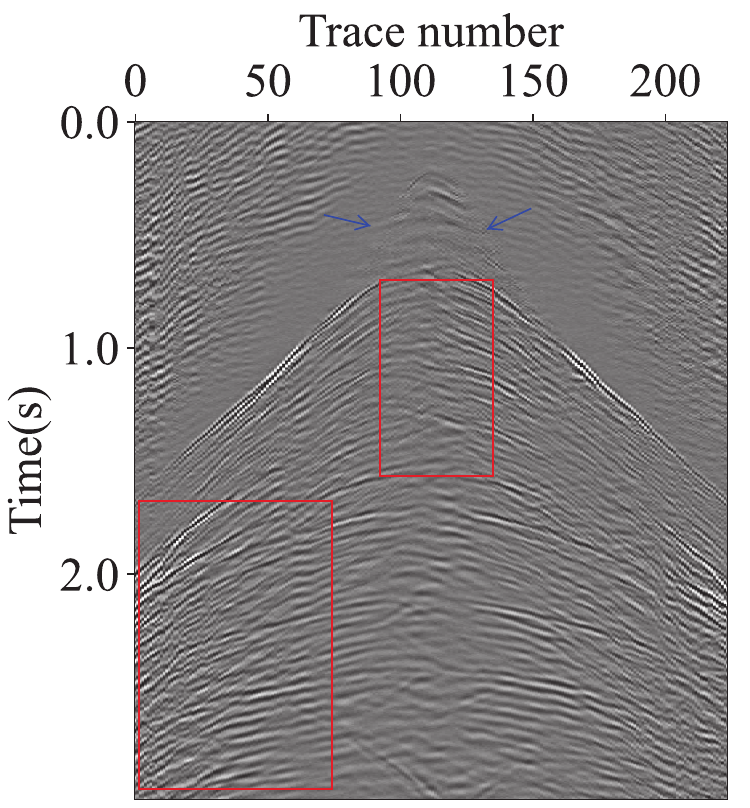}}
   \subfloat[\label{fig:lowf_line_b}]{
		\includegraphics[width=0.5\columnwidth]{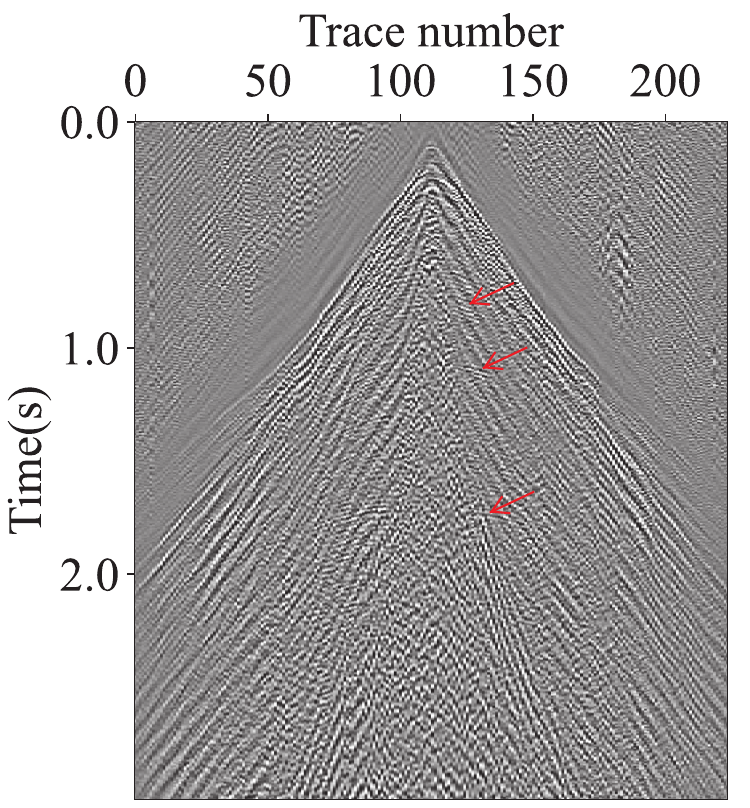}}	

  \caption{The clean data and ground roll predicted by the DDPM-2c with datasets2 for training. (a) The predicted clean data. (b) The predicted ground roll.}
    \label{fig:zerooffset_simple}
\end{figure}
\begin{figure} [htp!]
	\centering
 \subfloat[\label{fig:spec1_lowf_hr}]{
		\includegraphics[width=1\columnwidth]{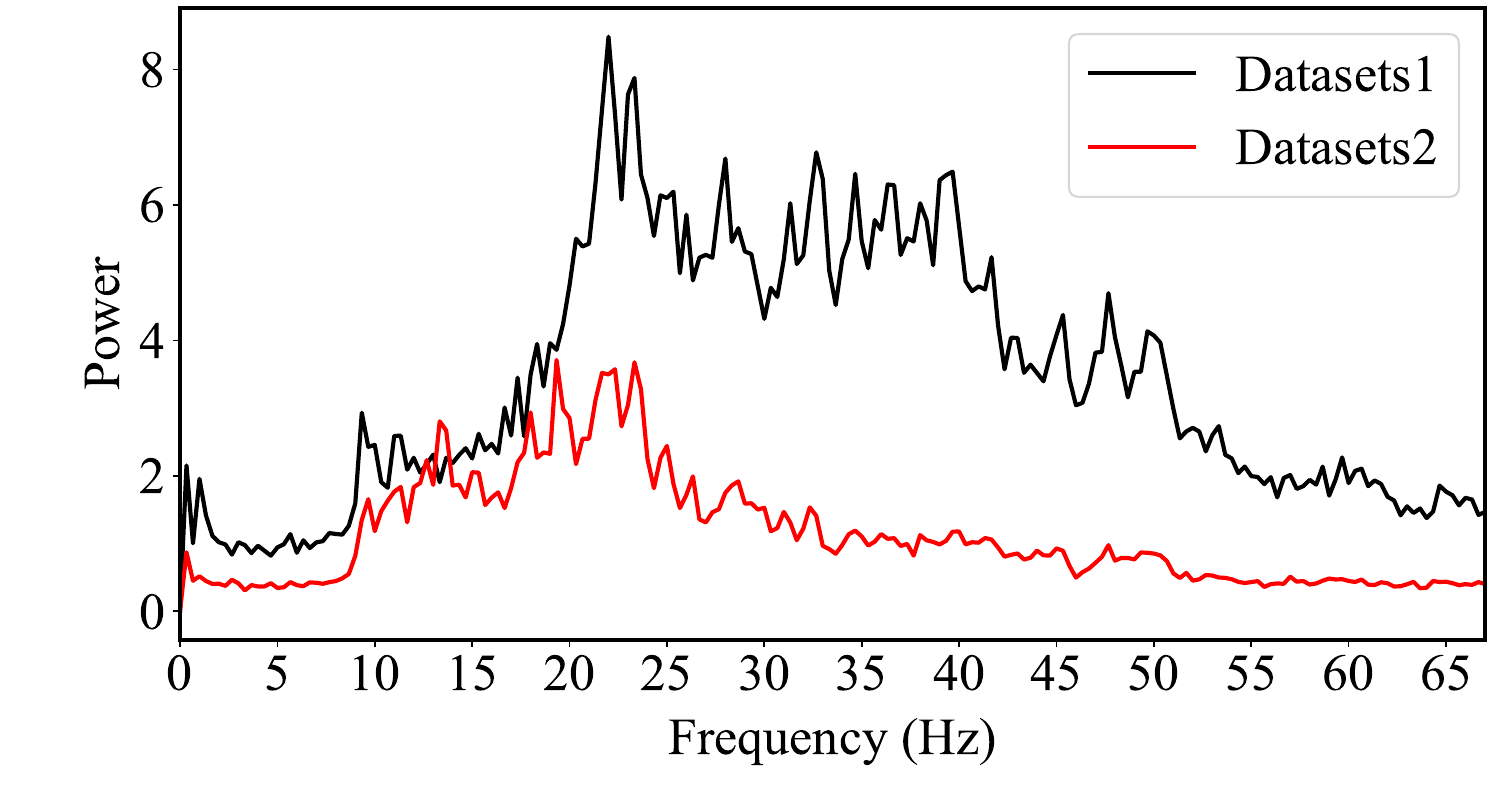}}\\
   \subfloat[\label{fig:spec1_lowf_lr}]{
		\includegraphics[width=1\columnwidth]{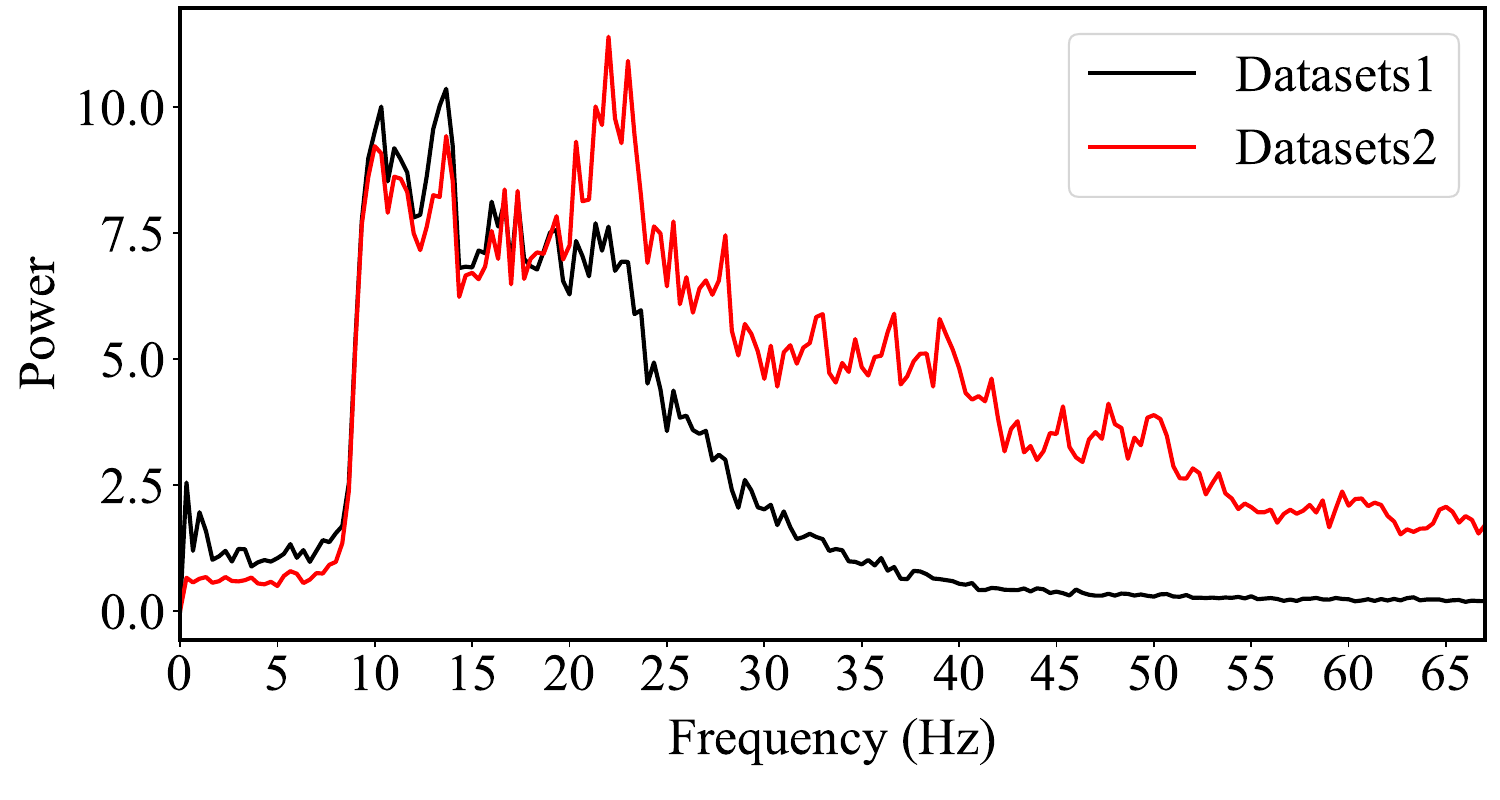}}	

  \caption{Amplitude spectra comparison of separated data using dataset1 and dataset2. (a) The separated clean data. (b) The separate ground roll data.}
    \label{fig:spec1_lowf_all}
\end{figure}

\subsection{DDIM}
We implemented a conventional sampling process of the DDPM in the c-DDPM. This sampling process is time-consuming because many timesteps are required. In order to accelerate the sampling, DDIM defines both the forward and reverse processes on a non-Markov chain instead of a Markov chain \cite{song2020denoising}. This non-Markov chain allows DDIM to sample within significantly reduced timesteps, making it much more computationally efficient. However, the reduced timesteps inevitably degrades the quality of the generation. Therefore, we employ the conventional sampling process of the DDPM instead of the sampling process of the DDIM. Here, we provide a brief illustration of the sampling process of the DDIM for reference. The sampling step in DDIM is
\begin{equation} \begin{split}
\mathbf{x}_{\tau_{i-1}}&=\sqrt{\alpha_{\tau_{i-1}}}\left(\frac{\mathbf{x}_{\tau_{i}}-\sqrt{1-\alpha_{\tau_{i}}}\epsilon_{\theta}(\mathbf{x}_{\tau_{i}},\tau_{i})}{\sqrt{\alpha_{\tau_{i}}}}\right)\\
&+\sqrt{1-\alpha_{\tau_{i-1}}-\sigma_{\tau_{i}}^{2}}\cdot\epsilon_{\theta}(\mathbf{x}_{\tau_{i}},\tau_{i})+\sigma_{\tau_{i}}\epsilon,
\end{split} \end{equation}
where $\tau_{i}$ is the sampled timesteps from the original range 0 ... $\tau$, $\alpha$ is the pre-defined noise schedule, and $\sigma_{\tau_{i}}^2=\eta\cdot\sqrt{(1-\alpha_{\tau_{i-1}})/(1-\alpha_{\tau_{i}})}\sqrt{(1-\alpha_{\tau_{i}}/\alpha_{\tau_{i-1}})}$. When $\eta=0$, there is no random noise added in every sampling step, thus the sampling process is fixed.  When $\eta=1$,  the sampling process is the same to the DDPM, only with reduced timesteps.
 \section{Conclusion}
 We propose to apply the c-DDPM to attenuate the ground-roll noise in the seismic recordings. We use the finite-difference modelling and convolution modelling methods to prepare the training datasets. After the training process, the c-DDPM can generate the clean data given the seismic recordings as condition. To improve the accuracy of prediction for ground roll, we further improve the c-DDPM to simultaneously generating the clean data and ground-roll data. Tests on one synthetic data and one field data show that the proposed method (the conventional c-DDPM and improved c-DDPM) performs better in ground-roll attenuation than the other methods using LTF and U-Net. Besides, the improved c-DDPM directly predict the ground roll with high accuracy. 
\ifCLASSOPTIONcaptionsoff
  \newpage
\fi
 \section{Appendix A} \label{appendix}
In the appendix, we will provide a brief introduction to the algorithms of MAE, MSE, SSIM and PSNR.
The MSE is calculated by 
\begin{equation}
\begin{aligned}
\mathrm{MAE=\frac1{mn}\sum_{j=1}^n\sum_{i=1}^m\left|\mathbf{y}_{i, j}-\hat{\mathbf{y}}_{i, j}\right|},
\end{aligned}
\end{equation}
where $m$ and $n$ is the size of the data, $y_{i, j}$ is the true data, and $\hat{y}_{i, j}$ is the estimated data.
The MSE is calculated by 
\begin{equation}
\begin{aligned}
\mathrm{MSE=\frac1{mn}\sum_{j=1}^n\sum_{i=1}^m(\mathbf{y}_{i, j}-\hat{\mathbf{y}}_{i, j})^2}.
\end{aligned}
\end{equation}
The PSNR is calculated by
\begin{equation}
\begin{aligned}
\mathrm{PSNR}=10\cdot\log_{10}\biggl(\frac{(2^n-1)^2}{\mathbf{MSE}}\biggr).
\end{aligned}
\end{equation}
The SSIM is calculated by
\begin{equation}
\begin{aligned}
\mathrm{SSIM}(\mathbf{y},\hat{\mathbf{y}})=\frac{(2\mu_{\mathbf{y}}\mu_{\hat{\mathbf{y}}}+c_1)(2\sigma_{\mathbf{y}\hat{\mathbf{y}}}+c_2)}{(\mu_{\mathbf{y}}^2+\mu_{\hat{\mathbf{y}}}^2+c_1)(\sigma_{\mathbf{y}}^2+\sigma_{\hat{\mathbf{y}}}^2+c_2)},
\end{aligned}
\end{equation}

where $\mu_{\mathbf{y}}, \mu_{\hat{\mathbf{y}}}$ is the mean of the true data $\mathbf{y}$ and estimated data $\hat{\mathbf{y}}$, respectively, $\sigma_{\mathbf{y}},\sigma_{\hat{\mathbf{y}}}$ is the variance of  $\mathbf{y}$ and $\hat{\mathbf{y}}$, respectively, $\sigma_{\mathbf{y}\hat{\mathbf{y}}}$ is the covariance of $\mathbf{y}$ and $\hat{\mathbf{y}}$, $c_1=(k_1L)^2,c_2=(k_2L)^2$, where $L$  is the dynamic scope, $k_1=0.01, k_2=0.03.$

\bibliographystyle{IEEEtran}
\bibliography{ref}

%

\newpage
\end{document}